\documentclass[aps,prab,twocolumn,nofootinbib,floatfix,longbibliography]{revtex4-2}
\usepackage{amsmath,amssymb,bm}
\usepackage{graphicx}
\usepackage{xcolor}
\definecolor{linkblue}{rgb}{0.10,0.10,0.55}
\usepackage[colorlinks=true,linkcolor=linkblue,citecolor=linkblue,urlcolor=linkblue]{hyperref}

\newcommand{\eq}[1]{(\ref{eq:#1})}
\newcommand{\T}{\mathsf{T}}

\newcommand{\Em}{\mathsf{E}}
\newcommand{\Id}{\mathsf{I}}
\newcommand{\ub}{\overline{\bm u}}
\newcommand{\Snu}{\Sigma_{\nu,j|z_0}}
\newcommand{\Jzz}{\mathcal{J}_{zz}}

\newcommand{\dd}{\mathrm{d}}

\begin{document}

\title{Intrabeam scattering of chirped electron bunches with non-Gaussian longitudinal distributions and slice mismatch}
\date{English draft, V7.1, 8 October 2026}

\author{Xiuji Chen}
\affiliation{Shanghai Advanced Research Institute, Chinese Academy of Sciences, Shanghai 201210, China}

\author{Wencai Cheng}
\email{\quad W Cheng, chengwc@sari.ac.cn}
\affiliation{Shanghai Advanced Research Institute, Chinese Academy of Sciences, Shanghai 201210, China}

\author{Si Chen}
\affiliation{Shanghai Advanced Research Institute, Chinese Academy of Sciences, Shanghai 201210, China}

\author{Duan Gu}
\email{\quad D Gu, gud@sari.ac.cn}
\affiliation{Shanghai Advanced Research Institute, Chinese Academy of Sciences, Shanghai 201210, China}
\author{Haixiao Deng}
\email{\quad HX Deng, denghx@sari.ac.cn}
\affiliation{Shanghai Advanced Research Institute, Chinese Academy of Sciences, Shanghai 201210, China}

\date{\today}

\begin{abstract}

Intrabeam scattering (IBS) can significantly increase the uncorrelated energy spread of high-brightness electron bunches in linear accelerators. For chirped bunches transported through beamlines with dispersion, however, the commonly used simulation codes elegant and RF-Track give substantially different predictions. To investigate this discrepancy, a semi-analytical slice model is developed for bunches with non-Gaussian longitudinal distributions and slice-dependent transverse mismatch. The model predicts slice energy-spread evolution in beamlines with dispersion and provides an independent reference for assessing the simulation results. Benchmarks in an achromatic arc show good agreement with RF-Track for both ideal Gaussian and realistic mismatched bunches in the tested cases. The proposed model provides a practical tool for evaluating IBS in linacs beyond the ideal-Gaussian approximation. The comparison also suggests directions for improving the accuracy and efficiency of PIC-based IBS simulations.

\end{abstract}

\maketitle

\section{Introduction}\label{sec:intro}

High-brightness electron beams are an essential basis for high-performance circular and linear accelerators.
Beam quality depends not only on the current and the transverse emittance but also on the uncorrelated energy spread and its variation along the bunch. In particular, for processes such as free-electron lasers (FELs), a large uncorrelated energy spread significantly degrades the quality of the emitted radiation~\cite{Bonifacio1984,HuangKim2007,Pellegrini2016}. Intrabeam scattering (IBS) arises from Coulomb interactions among particles of the same bunch; in the commonly used description of multiple small-angle scattering, it redistributes random momenta among different directions. For high-brightness electron beams that are transversely hot and longitudinally cold, IBS can appear mainly as growth of the uncorrelated energy spread (UES), while the transverse emittance changes comparatively little~\cite{Huang2002}. Research on IBS has long focused on its effects in storage rings~\cite{KuboOide2001,Bane2002ATF,Wolski2026}, but subsequent studies have shown that IBS effects also become increasingly evident in linear accelerators~\cite{DiMitri2020,Prat2022,Lucas2026,Cheng2026}.

The classical theory of IBS in storage rings was developed by Piwinski~\cite{Piwinski1974} and by Bjorken and Mtingwa~\cite{BM1983}. Coupled beams were treated by Kubo and Oide with a general Gaussian distribution and the beam envelope matrix~\cite{KuboOide2001}; Bane~\cite{Bane2002} and Kubo \textit{et al.}~\cite{KuboMtingwaWolski2005} simplified the growth-rate formulas in the high-energy limit, and Nagaitsev accelerated their numerical evaluation with Carlson symmetric elliptic integrals~\cite{Nagaitsev2005}. These theories relate the microscopic scattering process to the beam distribution and the lattice parameters, and provide growth rates for the evolution of the emittances and the UES. Unlike the equilibrium approximation used for storage rings, bunches in linear accelerators have different properties, such as a longitudinal profile far from Gaussian~\cite{XiaoBorland2009}, an energy chirp~\cite{XiaoBorland2009,Tsai2020,Pan2025}, and a transverse mismatch that varies from slice to slice, all of which can lead to IBS results different from those of a 6D Gaussian bunch. 

IBS in linacs has been widely studied. Based on the longitudinally cold condition and a transversely round-beam approximation, Huang estimated the IBS-induced energy-spread growth in an X-ray FEL driver~\cite{Huang2002}. IBS has also been studied extensively in less conventional linac beamlines and schemes, such as low-$\beta$ FODO sections~\cite{DiMitri2014}, recirculating beamlines and energy-recovery linacs~\cite{DiMitri2014,Xiao2008}, different FEL operation modes~\cite{Stupakov2011,Zhou2016,DiMitri2020}, and chicanes~\cite{Penn2014}. The interplay between IBS and the microbunching instability in linacs has also been investigated~\cite{Tsai2020,PerosaDiMitri2021}.

In general, IBS calculations for linac bunches, especially non-ideal bunches far from a Gaussian distribution, follow two approaches. The first is particle-in-cell (PIC) tracking with Monte Carlo collisions, as implemented in REPTIL~\cite{Gjonaj2022,Lucas2026} and RF-Track~\cite{DesireValdor2024,Desire2025}, which acts directly on macroparticles: the non-ideal IBS kick is retained at every step, and the actual evolution of the distribution is passed on to the next step instead of a frozen 6D Gaussian distribution. Converged results, however, require sufficient particle numbers, spatial resolution, and small time steps, which makes these codes costly for repeated use in parameter scans and beamline optimization. The second is the slice method, which is widely used in the dynamics of non-ideal bunches, for example in the multi-slice envelope description of space-charge-dominated photoinjector beams, in which the projected emittance is governed by the relative mismatch of the slice envelopes~\cite{Serafini1997,Ferrario2000}. Slice-based IBS models have likewise been studied extensively~\cite{Xiao2008,XiaoBorland2009,Lucas2026}. Compared with the PIC approach, the slice model is much faster. The slice model for IBS, however, has potential problems. After slicing, a conventional slice model evaluates the IBS of each slice with analytical formulas such as the methods in~\cite{Huang2002} or ~\cite{Lucas2026}. In a dispersive beamline, however, particles shift longitudinally and move between slices, which degrades the accuracy of such slice models, especially for beams with an energy chirp. The slice model in \textsc{elegant} provides a more general approach~\cite{Borland2000,XiaoBorland2009}: the bunch is divided into slices along the current longitudinal position, each slice is treated as Gaussian in the transverse coordinates and energy, and IBS is applied slice by slice. Although the initial transverse distribution can reasonably be assumed Gaussian, a slice that mixes particles from different parts of the bunch in a beamline with dispersion may be far from Gaussian. To summarize, RF-Track and \textsc{elegant} are two commonly used tools for IBS in linacs, and both have been benchmarked~\cite{Tsai2020,Desire2025}; however, to the authors' knowledge, these benchmarks do not include a bunch with a large energy chirp in a beamline with dispersion. Based on these considerations, we propose a new slice model for IBS in dispersive transport. Assuming an ideal Gaussian transverse distribution, the model follows the evolution of bunches with non-ideal longitudinal distributions under linear transport and accounts for the exchange of particles between slices in dispersive sections. Instead of replacing each slice with a Gaussian distribution of the same second moments, it evaluates the local collisions directly from the particles that meet at the same position, so that slices with different mean energies and transverse properties contribute with their actual relative motion.

\section{A General Collision integral for particles from different slices}\label{sec:I}

In the rest frame of a high-brightness electron bunch, the random motion is usually much colder in the longitudinal direction than in the transverse directions~\cite{Huang2002}. Coulomb collisions then transfer the energy of the random motion predominantly from the transverse to the longitudinal degree of freedom, so that the dominant IBS effect is the growth of the UES, while the transverse emittances change comparatively little~\cite{Huang2002}. All velocities are expressed in the bunch rest frame, in which the random motion is nonrelativistic. For two particles with velocities $\bm v_1^*$ and $\bm v_2^*$, the relative velocity and the kernel of the net longitudinal second moment are
\begin{equation}
\bm u=\bm v_1^*-\bm v_2^*,\qquad u=\|\bm u\|,\qquad
K_{zz}(\bm u)=\frac{u^2-3u_z^2}{u^3},
\label{eq:Kzz}
\end{equation}
Throughout this paper, the subscripts $x$, $y$, and $z$ denote the components along the corresponding directions, and bold symbols denote vectors. And the asterisk marks quantities in the bunch rest frame. The kernel $K_{zz}$ contains the combined contribution of friction and diffusion to the beam~\cite{KuboOide2001}.\footnote{Although the description above is motivated by the longitudinally cold condition, we do not take the cold limit $K_{zz}\to1/u$: the UES generated by IBS feeds back into the subsequent IBS, and particles from different slices that meet at the same position can have a nonzero mean longitudinal relative velocity, so the finite $u_z$ is retained in Eq.~\eq{Kzz}.} Let $f_j(\bm r,\bm v^*;s)$ denote the phase-space distribution of the particles of slice $j$ at the path length $s$, where $\int f_j\,\dd^3\bm r\,\dd^3\bm v^*=1$. The distribution is defined on volume elements that are infinitesimal on the macroscopic scale of the beam distribution but still contain many particles. On this scale, IBS is a local interaction~\cite{KuboOide2001}, so that two particles interact only when they are at the same position, $\bm r_1=\bm r_2=\bm r$; the microscopic impact parameters are accounted for by the Coulomb logarithm. 

The collisions between particles of slices $i$ and $j$ are then described by the pair integral
\begin{equation}
A_{ij}(s)=\int\dd^3\bm r\,\dd^3\bm v_1^*\,\dd^3\bm v_2^*\,f_i(\bm r,\bm v_1^*;s)\,f_j(\bm r,\bm v_2^*;s)\,K_{zz}(\bm u).
\label{eq:1e}
\end{equation}
At a position $\bm r$, the normalized velocity distribution of slice $j$ is $f_j/\rho_j$, where $\rho_j(\bm r;s)=\int f_j\,\dd^3\bm v^*$ is the position density of slice $j$. With $q_i$ the absolute charge of slice $i$ and $e$ the elementary charge, slice $i$ contains $q_i/e$ particles, and the density of phase-space is $(q_i/e)f_i$. Let $\sigma_{\delta,j}(\bm r)$ denote the uncorrelated energy spread of the slice-$j$ particles at $\bm r$. Following Eqs.~(39)--(44) in Ref.~\cite{KuboOide2001}, the rate at which collisions with slice $i$ change $\sigma_{\delta,j}^2(\bm r)$ is
\begin{equation}
\begin{aligned}
\left(\frac{\dd\sigma_{\delta,j}^2(\bm r)}{\dd s}\right)_{\mathrm{IBS},i}
={}&\frac{4\pi r_e^2c\ln\Lambda}{\beta_0^3\gamma_0^2}\,\frac{q_i}{e\,\rho_j(\bm r)}\\
&\times\int f_j(\bm r,\bm v_1^*)\,f_i(\bm r,\bm v_2^*)\,K_{zz}(\bm u)\,\dd^3\bm v_1^*\,\dd^3\bm v_2^*,
\end{aligned}
\label{eq:2d}
\end{equation}
where $r_e$ is the classical electron radius, $\beta_0$ and $\gamma_0$ are the reference relativistic factors, and $\ln\Lambda$ is the Coulomb logarithm. By averaging Eq.~\eq{2d} over the positions of slice $j$ with the weight $\rho_j$ and summing over all slices $i$, the IBS-induced change of the energy spread of slice $j$ can be expressed as
\begin{equation}
S_{\delta,j}(s)=\left(\frac{\dd\sigma_{\delta,j}^2}{\dd s}\right)_{\mathrm{IBS}}=\frac{4\pi r_e^2c\ln\Lambda(s)}{e\,\beta_0^3\gamma_0^2}\sum_iq_iA_{ij}(s),
\label{eq:2}
\end{equation}
where the sum includes both $A_{jj}$ within slice $j$ and $A_{ij}$ between different slices. In the slice model, the bunch is divided into slices of equal width according to the initial longitudinal coordinate, and the particles are assumed to be uniformly distributed within each slice; after dispersive transport, different slices can overlap in space. The entrance interval of slice $j$ is
\begin{equation}
I_j=[\bar z_j-L/2,\ \bar z_j+L/2).
\label{eq:3}
\end{equation}
Here, $\bar z_j$ is the center of slice $j$ and $L$ is the slice width, while the charges, Twiss parameters, and emittances of the slices can differ. $f_j(\bm r,\bm v^*;s\mid z_0)$ denotes the distribution of slice $j$ with the initial longitudinal coordinate $z_0$. Then the distribution of slice $j$ can be expressed as
\begin{equation}
f_j(\bm r,\bm v^*;s)=\frac1L\int_{I_j}f_j(\bm r,\bm v^*;s\mid z_0)\,\dd z_0,
\label{eq:4}
\end{equation}
Substituting Eq.~\eq{4} into Eq.~\eq{1e}, the $A_{ij}$ can be expressed as
\begin{equation}
\begin{aligned}
A_{ij}(s)={}&\frac1{L^2}\int_{I_i}\dd z_{01}\int_{I_j}\dd z_{02}\int\dd^3\bm r\,\dd^3\bm v_1^*\,\dd^3\bm v_2^*\\
&\times f_i(\bm r,\bm v_1^*;s\mid z_{01})\,f_j(\bm r,\bm v_2^*;s\mid z_{02})\,K_{zz}(\bm u).
\end{aligned}
\label{eq:5}
\end{equation}
$z_{01}$ and $z_{02}$ are the initial longitudinal coordinates of the two particles, and $z$ is the longitudinal component of their current common position $\bm r$. Because the transport contains transverse--longitudinal coupling and dispersion, a fixed $z_0$ generally does not fix the current $z$; the average over the entrance coordinates therefore cannot be identified with the integral over the common position.

\begin{figure}
    \centering
    \includegraphics[width=1\linewidth]{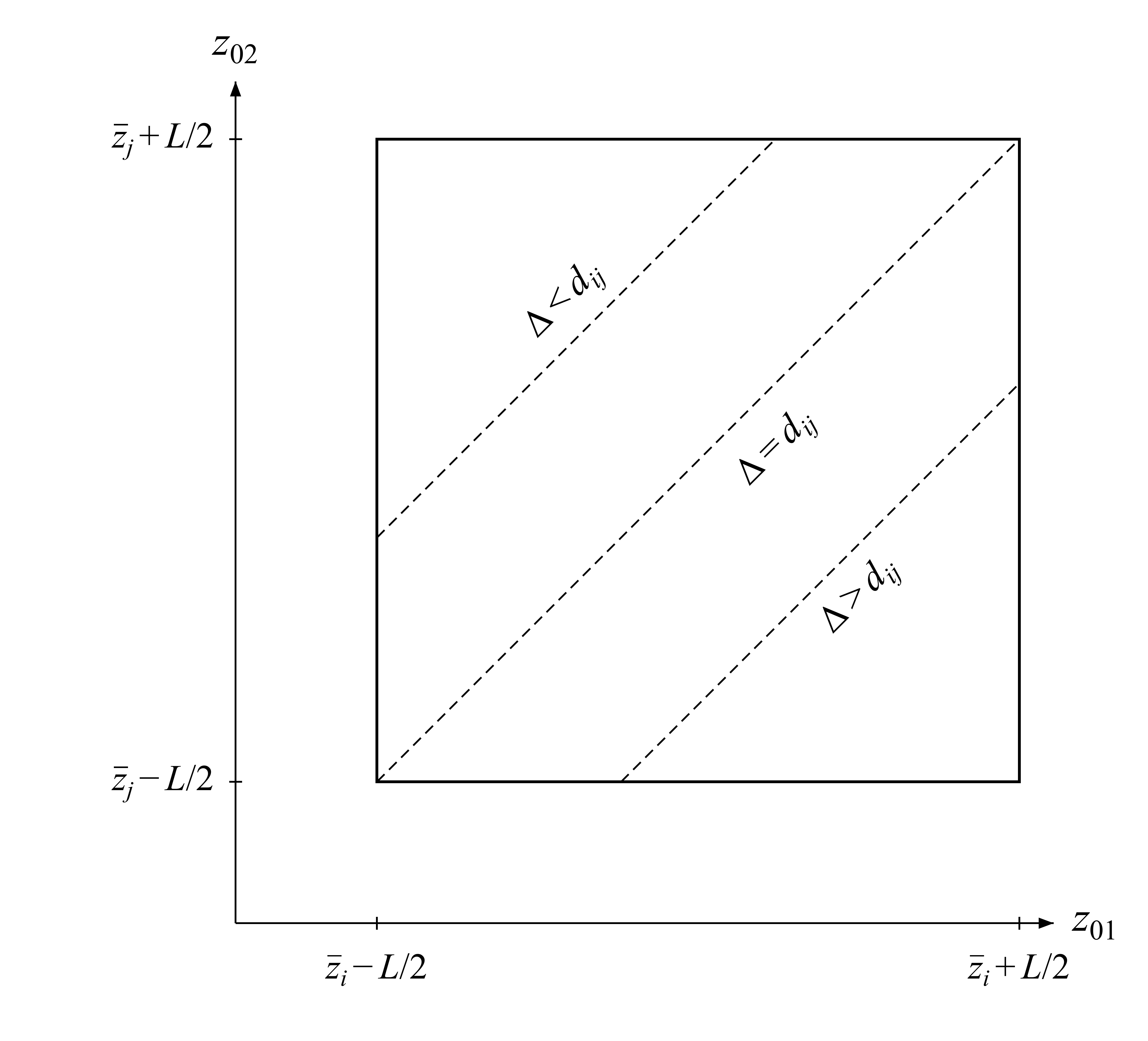}
    \caption{Pairing region of the slices $i$ and $j$. Each point in the square $I_i\times I_j$ represents a pair of $(z_{01},z_{02})$. The dashed lines represent $\Delta=z_{01}-z_{02}$.}
    \label{fig:z_density}
\end{figure}

The entrance-coordinate difference and the separation between the slice centers are defined as $\Delta=z_{01}-z_{02},\quad d_{ij}=\bar z_i-\bar z_j$. As shown in Fig.~\ref{fig:z_density}, the $(z_{01},z_{02})$ of a particle pair are uniformly distributed over the square $I_i\times I_j$ with the density $1/L^2$. Each value of $\Delta$ corresponds to a line $z_{01}-z_{02}=\Delta$ parallel to the diagonal of the square, as illustrated by the dashed lines in Fig.~\ref{fig:z_density}. The probability that the difference lies between $\Delta$ and $\Delta+\dd\Delta$ is therefore the area of the thin strip around this line inside the square, divided by $L^2$. This area is largest for the diagonal $\Delta=d_{ij}$ and decreases linearly on both sides until it vanishes at $\Delta=d_{ij}\pm L$, so that the normalized density of $\Delta$ is the triangular weight density:
\begin{equation}
\omega_{ij}(\Delta)=\frac{[L-|\Delta-d_{ij}|]_+}{L^2}.
\label{eq:5b}
\end{equation}
Here, $[a]_+=\max(a,0)$. And $\omega_{ij}(\Delta)$ is supported on $[d_{ij}-L,d_{ij}+L]$. With this weight, the double average over $z_{01}$ and $z_{02}$ reduces to a single integral over $\Delta$: for any $F(\Delta)$, we have
\begin{equation}
\frac1{L^2}\int_{I_i}\dd z_{01}\int_{I_j}\dd z_{02}\,F(\Delta)=\int_{d_{ij}-L}^{d_{ij}+L}\omega_{ij}(\Delta)F(\Delta)\,\dd\Delta.
\label{eq:5c}
\end{equation}
The collision term of a beam distribution is built from the product of the two phase-space densities at the same position~\cite{KuboOide2001}. Therefore, for fixed $z_{01}$ and $z_{02}$, the integral over $\bm r$, $\bm v_1^*$, and $\bm v_2^*$ in Eq.~\eq{5} can be separated into a position part and a velocity part. Each distribution is written as $f_j=\rho_jg_j$, where $\rho_j(\bm r;s\mid z_0)$ is the position density and $g_j=f_j/\rho_j$ is the normalized local velocity distribution at $\bm r$. Then, $A_{ij}$ can be expressed as
\begin{equation}
A_{ij}(s)=\frac1{L^2}\int_{I_i}\dd z_{01}\int_{I_j}\dd z_{02}\,P_{ij}\,\Jzz^{ij},
\label{eq:7}
\end{equation}
where
\begin{equation}
\begin{aligned}
P_{ij}(z_{01},z_{02};s)&=\int\rho_i(\bm r;s\mid z_{01})\,\rho_j(\bm r;s\mid z_{02})\,\dd^3\bm r,\\
\Jzz^{ij}(z_{01},z_{02};s)&=\int p_{u,ij}(\bm u;z_{01},z_{02},s)\,K_{zz}(\bm u)\,\dd^3\bm u.
\end{aligned}
\label{eq:8}
\end{equation}
Here, $P_{ij}$ is the position overlap factor. It is the probability density for the particles of slices $i$ and $j$ to meet at the same position. $p_{u,ij}$ is the local relative-velocity distribution of the particle pairs that meet at the same position, normalized by $P_{ij}$. $\Jzz^{ij}$ is the velocity factor, i.e., the average of $K_{zz}$ over $p_{u,ij}$. Equations~\eq{7} and \eq{8} give a general form of the IBS collision integral, which holds for arbitrary distributions, such as the cross terms between the two components of a bi-Gaussian beam~\cite{Parzen2004}. When $P_{ij}$ and $\Jzz^{ij}$ depend on $z_{01}$ and $z_{02}$ through $\Delta$, their product can be taken as $F(\Delta)$ in Eq.~\eq{5c}; hence, $A_{ij}$ becomes
\begin{equation}
A_{ij}(s)=\int_{d_{ij}-L}^{d_{ij}+L}\omega_{ij}(\Delta)\,P_{ij}(\Delta;s)\,\Jzz^{ij}(\Delta;s)\,\dd\Delta.
\label{eq:10}
\end{equation}
In the above discussion, a general form of $A_{ij}$ is given formally; this structure holds for arbitrary distributions. Substituting into Eq.~\eq{2}, the variance in energy spread caused by IBS can be obtained for arbitrary distributions. In the following, a more specific form is derived by assuming a Gaussian transverse distribution in each slice, which usually deviates much less from a Gaussian than the longitudinal distribution, together with a linear energy chirp; under these assumptions, the condition for Eq.~\eq{10} is satisfied.

\section{Transport and position--momentum decorrelation of an energy-chirped beam in an achromat}\label{sec:II}

In this study, the transverse distribution of each slice is approximated as Gaussian, and the initial transverse offsets are neglected. The energy chirp is approximated as linear, $\delta_0=hz_0+\eta_0$, where $h$ is the linear chirp common to the whole bunch and $\eta_0$ is the uncorrelated energy deviation within the slice. Then, $\eta_0$ in each slice follows a Gaussian distribution with the variance $\sigma_{\eta,j,0}^2$. To evaluate $P_{ij}$ and $p_{u,ij}$ in Eq.~\eq{8}, the phase-space coordinates and the reference momentum are
\begin{equation}
\bm X=(x,x',y,y',z,\delta)^{\T},\qquad \delta=\frac{p-p_0}{p_0},\qquad p_0=\beta_0\gamma_0m_ec.
\label{eq:11}
\end{equation}
Here, $p_0$ is the reference momentum and $m_e$ is the electron mass. For paraxial motion and nonrelativistic random motion in the bunch frame, the velocity difference can be expressed as
\begin{equation}
\bm u=T_v(\bm P_1-\bm P_2),\qquad \bm P=(x',y',\delta)^{\T},
\label{eq:12}
\end{equation}
where
\begin{equation}
T_v=\begin{pmatrix}
\beta_0\gamma_0 c & 0 & 0\\
0 & \beta_0\gamma_0 c & 0\\
0 & 0 & \beta_0 c
\end{pmatrix}.
\end{equation}
Moreover, the transverse second moments of slice $j$ at the entrance can be expressed with the Twiss parameters as
\begin{equation}
\Sigma^{(0)}_{vv',j}=\varepsilon_{v,j,0}\begin{pmatrix}\beta_{v,j,0}&-\alpha_{v,j,0}\\-\alpha_{v,j,0}&\gamma_{v,j,0}\end{pmatrix},\qquad v=x,y,
\label{eq:14}
\end{equation}
where $\varepsilon_{v,j,0}$ is the geometric emittance, and $\beta_{v,j,0}$, $\alpha_{v,j,0}$, and $\gamma_{v,j,0}=(1+\alpha_{v,j,0}^2)/\beta_{v,j,0}$ are the Twiss functions. The longitudinal second moments are
\begin{equation}
\Sigma^{(0)}_{z\delta,j}=\begin{pmatrix}L^2/12&hL^2/12\\hL^2/12&\sigma^2_{\eta,j,0}+h^2L^2/12\end{pmatrix},
\label{eq:16}
\end{equation}
where $L^2/12$ is the variance of $z_0$ uniformly distributed over the slice width, and the terms proportional to $h$ come from the linear chirp. Thus, the second-moment matrix of slice $j$ at the entrance is
\begin{equation}
\Sigma_{X,j}(0)=\begin{pmatrix}
\Sigma^{(0)}_{xx',j}&0&0\\
0&\Sigma^{(0)}_{yy',j}&0\\
0&0&\Sigma^{(0)}_{z\delta,j}
\end{pmatrix},
\label{eq:17}
\end{equation}
Correspondingly, the distribution of slice $j$ at the entrance is
\begin{equation}
\begin{aligned}
f_{0j}(\bm X_0)={}&\frac{{\mathcal{I}_j}(z_0)}{L}\,G_2\bigl((x_0,x_0')^{\T};\Sigma^{(0)}_{xx',j}\bigr)\,G_2\bigl((y_0,y_0')^{\T};\Sigma^{(0)}_{yy',j}\bigr)\\
&\times G_1\bigl(\delta_0-hz_0;\sigma^2_{\eta,j,0}\bigr),
\end{aligned}
\label{eq:15}
\end{equation}
where $\mathcal{I}_j(z_0)$ equals $1$ for $z_0\in I_j$ and $0$ otherwise, and
\[
G_d(\bm\xi;\Sigma)=\frac{1}{(2\pi)^{d/2}\sqrt{\det\Sigma}}\exp\Bigl(-\frac12\bm\xi^{\T}\Sigma^{-1}\bm\xi\Bigr)
\]
is the $d$-dimensional normalized Gaussian density with zero mean and covariance $\Sigma$. The three factors in Eq.~\eq{15} describe the uniform distribution of $z_0$ within the slice, the Gaussian transverse distributions, and the Gaussian uncorrelated energy deviation around the linear chirp, respectively.

Since the deflection occurs only in the horizontal plane, the transfer matrix from the entrance to any position $s$ is
\begin{equation}
R(s)=\begin{pmatrix}
R_{11}&R_{12}&0&0&0&R_{16}\\
R_{21}&R_{22}&0&0&0&R_{26}\\
0&0&R_{33}&R_{34}&0&0\\
0&0&R_{43}&R_{44}&0&0\\
R_{51}&R_{52}&0&0&R_{55}&R_{56}\\
0&0&0&0&0&1
\end{pmatrix}.
\label{eq:19}
\end{equation}
Here, the longitudinal position of a particle is $z=\kappa z_0+R_{51}x_0+R_{52}x_0'+R_{56}\eta_0$ with $\kappa=R_{55}+hR_{56}$. Therefore, the local overlap of each slice with other slices is determined jointly by the transverse and longitudinal coordinates. At the entrance, the centroid of the particles of slice $j$ with a given $z_0$ is $\bm d(0)z_0$, where $\bm d(0)=(0,0,0,0,1,h)^{\T}$. Then, at the position $s$, the centroid can be expressed as
\begin{equation}
\begin{aligned}
\langle\bm X(s)\rangle_{j,z_0}&=\bm d(s)z_0,\\
\bm d(s)&=R(s)\bm d(0)=(hR_{16},hR_{26},0,0,\kappa,h)^{\T}.
\end{aligned}
\label{eq:21a}
\end{equation}
For $P_{ij}$ and $p_{u,ij}$, only the motion relative to the centroid $\bm d(s)z_0$ is needed; the deviation from the centroid and its second-moment matrix are defined as
\begin{equation}
\bm\nu_j(s\mid z_0)=\bm X(s)-\bm d(s)z_0,\qquad
\Snu(s)=\bigl\langle\bm\nu_j\bm\nu_j^{\T}\bigr\rangle_{j,z_0}.
\label{eq:21b}
\end{equation}
Here, $\Snu$ is the second-moment matrix of the particles of slice $j$ with a given $z_0$, which describes the uncorrelated motion around the centroid. Within a slice, the centered distribution is independent of $z_0$, so that $\Snu$ does not depend on $z_0$. Because $z_0$ is uniformly distributed over the slice width, the second-moment matrix of the whole slice and $\Snu$ satisfy
\begin{equation}
\Snu(s)=\Sigma_{X,j}(s)-\frac{L^2}{12}\bm d(s)\bm d^{\T}(s).
\label{eq:22}
\end{equation}

\subsection{\texorpdfstring{The position overlap $P_{ij}$}{The position overlap Pij}}
At the entrance, Eq.~\eq{22} gives the longitudinal block of $\Snu$ as $\begin{pmatrix}0&0\\0&\sigma^2_{\eta,j,0}\end{pmatrix}$, while its transverse blocks remain those of Eq.~\eq{14}. Here, $[\Snu]_{55}=0$ because the particles with the same $z_0$ have the same longitudinal position. And $[\Snu]_{55}$ becomes nonzero through $R_{5i}$ ($i=1,2,6$) in the beamline with dispersion. Different from the second-moment matrix $\Sigma_{X,j}$ of the whole slice in Eq.~\eq{17}, $\Snu$ describes the motion of the particles with the same $z_0$ around their centroid, i.e., the local uncorrelated motion.

In this study, the coupling with the vertical plane is neglected.
Thus, the position block of $\Snu$ is
\begin{equation}
\Sigma_{\nu r,j}=\begin{pmatrix}
[\Snu]_{11}&0&[\Snu]_{15}\\
0&[\Snu]_{33}&0\\
{}[\Snu]_{15}&0&[\Snu]_{55}
\end{pmatrix}.
\label{eq:S6}
\end{equation}
For the particles of slice $j$ with a given $z_0$, the position density can then be expressed as
\begin{equation}
\rho_j(\bm r;s\mid z_0)=G_3\bigl(\bm r-\bm d_rz_0;\Sigma_{\nu r,j}\bigr),
\label{eq:S7}
\end{equation}
where $\bm d_r=(hR_{16},0,\kappa)^{\T}$ is the position part of $\bm d(s)$.

Consider two particles from slices $i$ and $j$ with the initial longitudinal coordinates $z_{01}$ and $z_{02}$, respectively.
Since their positions are independent variables, the position difference $\bm\ell=\bm r_1-\bm r_2$ is also Gaussian. Its mean is the difference of the two centroids, $\bm d_rz_{01}-\bm d_rz_{02}=\bm d_r\Delta$, and its covariance is the sum of the two covariances,
\begin{equation}
\begin{aligned}
p_\ell(\bm\ell)&=\int\rho_i(\bm r+\bm\ell;s\mid z_{01})\,\rho_j(\bm r;s\mid z_{02})\,\dd^3\bm r\\
&=G_3\bigl(\bm\ell-\bm d_r\Delta;\Sigma_{\nu r,ij}\bigr).
\end{aligned}
\label{eq:pl}
\end{equation}
where $\Sigma_{\nu r,ij}$ is the covariance of $\bm\ell$. Writing the positions as $\bm r_1=\bm d_rz_{01}+\bm\nu_{r,i}$ and $\bm r_2=\bm d_rz_{02}+\bm\nu_{r,j}$, where $\bm\nu_{r,i}$ and $\bm\nu_{r,j}$ are the position parts of the deviations in Eq.~\eq{21b}, the position difference becomes $\bm\ell=\bm d_r\Delta+\bm\nu_{r,i}-\bm\nu_{r,j}$. Since the two particles are independent, $\langle\bm\nu_{r,i}\bm\nu_{r,j}^{\T}\rangle=0$. Thus,
\begin{equation}
\Sigma_{\nu r,ij}=\bigl\langle(\bm\nu_{r,i}-\bm\nu_{r,j})(\bm\nu_{r,i}-\bm\nu_{r,j})^{\T}\bigr\rangle=\Sigma_{\nu r,i}+\Sigma_{\nu r,j}.
\label{eq:Sigrij}
\end{equation}
Since each slice keeps its own $\Snu$, the mismatch between slices is retained in $\Sigma_{\nu r,ij}$. For the case that $\bm\ell=0$, $P_{ij}$ in Eq.~\eq{8} can be expressed as
\begin{equation}
P_{ij}(\Delta;s)=p_\ell(0)=G_3\bigl(\bm d_r\Delta;\Sigma_{\nu r,ij}\bigr),
\label{eq:28b}
\end{equation}
which is the probability density for the two particles to be at the same position. Since $\Sigma_{\nu r,i}$ and $\Sigma_{\nu r,j}$ do not depend on $z_0$, $P_{ij}$ depends on $z_{01}$ and $z_{02}$ only through $\Delta$, as required by Eq.~\eq{10}. And $P_{ij}$ can also be expressed as
\begin{equation}
P_{ij}(\Delta;s)=C_{ij}\exp\Bigl(-\frac{\alpha_{ij}\Delta^2}{2}\Bigr),
\label{eq:32}
\end{equation}
where
\begin{equation}
\alpha_{ij}=\bm d_r^{\T}\Sigma_{\nu r,ij}^{-1}\bm d_r,\qquad
C_{ij}=\frac{1}{(2\pi)^{3/2}\sqrt{\det\Sigma_{\nu r,ij}}}.
\label{eq:alphaC}
\end{equation}
\subsection{\texorpdfstring{The local relative-velocity distribution $p_{u,ij}$}{The local relative-velocity distribution pu}}

Similar to the positions $\bm r_1$ and $\bm r_2$, the momenta of the two particles can be written as $\bm P_1=\bm d_Pz_{01}+\bm\nu_{P,i}$ and $\bm P_2=\bm d_Pz_{02}+\bm\nu_{P,j}$, where $\bm d_P=(hR_{26},0,h)^{\T}$ is the momentum part of $\bm d(s)$ and $\bm\nu_{P,i}$ and $\bm\nu_{P,j}$ are the momentum parts of the deviations in Eq.~\eq{21b}. By substituting them into Eq.~\eq{12}, the relative velocity becomes
\begin{equation}
\bm u=T_v\bigl(\bm d_P\Delta+\bm\nu_{P,i}-\bm\nu_{P,j}\bigr).
\label{eq:26}
\end{equation}
Since the deviations of two different particles are uncorrelated, $\langle\bm\nu_{P,i}\bm\nu_{P,j}^{\T}\rangle=0$ and $\langle\bm\nu_{P,i}\bm\nu_{r,j}^{\T}\rangle=\langle\bm\nu_{P,j}\bm\nu_{r,i}^{\T}\rangle=0$. Thus,
\begin{equation}
\begin{aligned}
\Sigma_{\nu P,ij}&=\bigl\langle(\bm\nu_{P,i}-\bm\nu_{P,j})(\bm\nu_{P,i}-\bm\nu_{P,j})^{\T}\bigr\rangle=\Sigma_{\nu P,i}+\Sigma_{\nu P,j},\\
\Sigma_{\nu P\nu r,ij}&=\bigl\langle(\bm\nu_{P,i}-\bm\nu_{P,j})(\bm\nu_{r,i}-\bm\nu_{r,j})^{\T}\bigr\rangle=\Sigma_{\nu P\nu r,i}+\Sigma_{\nu P\nu r,j},
\end{aligned}
\label{eq:27}
\end{equation}
where the momentum block and the momentum--position block of $\Snu$ are
\begin{equation}
\begin{aligned}
\Sigma_{\nu P,j}&=\begin{pmatrix}
[\Snu]_{22}&0&[\Snu]_{26}\\
0&[\Snu]_{44}&0\\
{}[\Snu]_{26}&0&[\Snu]_{66}
\end{pmatrix},\\
\Sigma_{\nu P\nu r,j}&=\begin{pmatrix}
[\Snu]_{12}&0&[\Snu]_{25}\\
0&[\Snu]_{34}&0\\
{}[\Snu]_{16}&0&[\Snu]_{56}
\end{pmatrix}.
\end{aligned}
\label{eq:28}
\end{equation}

By its definition in Eq.~\eq{8}, $p_{u,ij}$ is the distribution of $\bm u$ for the pairs at the same position, normalized by $P_{ij}$ in Eq.~\eq{28b}.
At the same position, $\bm r_1-\bm r_2=\bm d_r\Delta+\bm\nu_{r,i}-\bm\nu_{r,j}=0$, so that
\begin{equation}
\bm\nu_{r,ij}\equiv\bm\nu_{r,i}-\bm\nu_{r,j}=-\bm d_r\Delta.
\label{eq:samepos}
\end{equation}
The momentum deviation $\bm\nu_{P,ij}\equiv\bm\nu_{P,i}-\bm\nu_{P,j}$ is written as
\begin{equation}
\bm\nu_{P,ij}=H_{ij}\bm\nu_{r,ij}+\bm\epsilon_{ij},\qquad
H_{ij}=\Sigma_{\nu P\nu r,ij}\Sigma_{\nu r,ij}^{-1},
\label{eq:S9}
\end{equation}
where $H_{ij}$ is the matrix form of the correlation coefficients between the momentum difference and the position difference, and $\bm\epsilon_{ij}$ is the residual uncorrelated with $\bm\nu_{r,ij}$.\footnote{Requiring $\langle\bm\epsilon_{ij}\bm\nu_{r,ij}^{\T}\rangle=\Sigma_{\nu P\nu r,ij}-H_{ij}\Sigma_{\nu r,ij}=0$ gives $H_{ij}$ in Eq.~\eq{S9}, and the covariance of $\bm\epsilon_{ij}$ is then $\Sigma_{\nu P,ij}-H_{ij}\Sigma_{\nu P\nu r,ij}^{\T}$. In one dimension, $H=\langle\nu_P\nu_r\rangle/\langle\nu_r^2\rangle$; familiar examples of this form are $-\alpha/\beta=\langle xx'\rangle/\langle x^2\rangle$ for a betatron phase space and the dispersion $D=\langle x\delta\rangle/\langle\delta^2\rangle$.}. By substituting Eq.~\eq{samepos}, the momentum difference at the same position becomes
\begin{equation}
\bm P_1-\bm P_2=\bm d_P\Delta+\bm\nu_{P,ij}=(\bm d_P-H_{ij}\bm d_r)\Delta+\bm\epsilon_{ij}.
\label{eq:30}
\end{equation}
Due to the same-position constraint, the position deviation within the slices is replaced by the centroid difference, which is a more convenient form for the calculation. Since $\bm\epsilon_{ij}$ and $\bm\nu_{r,ij}$ are Gaussian and uncorrelated, $\bm\epsilon_{ij}$ is independent of $\bm\nu_{r,ij}$; that is, the correlation with the position within the slices is removed. Therefore, at the same position, $\bm P_1-\bm P_2$ is Gaussian with the mean $(\bm d_P-H_{ij}\bm d_r)\Delta$ and the covariance of $\bm\epsilon_{ij}$. By substituting Eq.~\eq{30} into Eq.~\eq{26}, $p_{u,ij}$ can be expressed as
\begin{equation}
p_{u,ij}(\bm u;\Delta,s)=G_3\bigl(\bm u-\bm b_{ij}\Delta;\Sigma_{u,ij}\bigr),
\label{eq:38}
\end{equation}
where
\begin{equation}
\bm b_{ij}=T_v(\bm d_P-H_{ij}\bm d_r),\quad
\Sigma_{u,ij}=T_v\bigl(\Sigma_{\nu P,ij}-H_{ij}\Sigma_{\nu P\nu r,ij}^{\T}\bigr)T_v^{\T}.
\label{eq:33}
\end{equation}
Here, $\bm b_{ij}\Delta$ is the mean relative velocity of the particles meeting at the same position. It vanishes for the same slice at $\Delta=0$, whereas it is generally nonzero for $\Delta\neq0$ or for different slices, for example because of the energy chirp and dispersion. $\Sigma_{u,ij}$ is the local uncorrelated spread of the relative velocity, which does not depend on $\Delta$ and contains the mismatch between the two slices. Thus, both $P_{ij}$ and $\Jzz^{ij}$ depend on $z_{01}$ and $z_{02}$ through $\Delta$, so that Eq.~\eq{10} applies. By substituting Eq.~\eq{32} into Eq.~\eq{10}, $A_{ij}$ can be expressed as
\begin{equation}
A_{ij}(s)=C_{ij}\int_{d_{ij}-L}^{d_{ij}+L}\omega_{ij}(\Delta)\,e^{-\alpha_{ij}\Delta^2/2}\,\Jzz^{ij}(\Delta;s)\,\dd\Delta,
\label{eq:41}
\end{equation}
where, with Eq.~\eq{38}, the velocity factor is
\begin{equation}
\Jzz^{ij}(\Delta;s)=\int G_3\bigl(\bm u-\bm b_{ij}\Delta;\Sigma_{u,ij}\bigr)\,K_{zz}(\bm u)\,\dd^3\bm u.
\label{eq:40}
\end{equation}
In this section, $P_{ij}$ and $p_{u,ij}$ of any slice pair are obtained from the current $\Snu$ of the two slices, and the mismatch and the mean relative velocity between the slices are retained in $\Sigma_{u,ij}$ and $\bm b_{ij}\Delta$. In the following section, the velocity factor with a nonzero mean relative velocity and its average over $\Delta$ are discussed.

\section{Analytical reduction of the pair integral}\label{sec:III}

In this section, the velocity factor in Eq.~\eq{40} and the average over $\Delta$ in Eq.~\eq{41} are evaluated. Using the auxiliary-integral representation commonly used in IBS calculations~\cite{BM1983,KuboOide2001}, for $u>0$,
\begin{equation}
\frac1u=\frac1{\sqrt\pi}\int_0^\infty\tau^{-1/2}e^{-\tau u^2}\,\dd\tau,\qquad
\frac1{u^3}=\frac2{\sqrt\pi}\int_0^\infty\tau^{1/2}e^{-\tau u^2}\,\dd\tau.
\label{eq:42}
\end{equation}
By substituting Eq.~\eq{42} into Eq.~\eq{40}, the velocity factor can be expressed as
\begin{equation}
\Jzz^{ij}=\frac1{\sqrt\pi}\int_0^\infty\frac{\dd\tau}{\sqrt\tau}\Bigl[\bigl\langle e^{-\tau u^2}\bigr\rangle-6\tau\bigl\langle u_z^2e^{-\tau u^2}\bigr\rangle\Bigr],
\label{eq:43}
\end{equation}
where $\langle\cdot\rangle$ denotes the average over $p_{u,ij}$. 

With Eq.~\eq{38}, the first average in Eq.~\eq{43} can be written as
\begin{equation}
\bigl\langle e^{-\tau u^2}\bigr\rangle=\int G_3(\bm u-\ub;\Sigma_{u,ij})\,e^{-\tau u^2}\,\dd^3\bm u,
\label{eq:avgint}
\end{equation}
and the second average contains an additional factor $u_z^2$. With $\ub=\bm b_{ij}\Delta$, the product of the two exponentials gives the exponent
\begin{equation}
\begin{aligned}
&-\tfrac12(\bm u-\ub)^{\T}\Sigma_{u,ij}^{-1}(\bm u-\ub)-\tau\bm u^{\T}\bm u\\
&\quad=-\tfrac12\bm u^{\T}\bigl(\Sigma_{u,ij}^{-1}+2\tau\Id_3\bigr)\bm u+\bm u^{\T}\Sigma_{u,ij}^{-1}\ub-\tfrac12\ub^{\T}\Sigma_{u,ij}^{-1}\ub,
\end{aligned}
\label{eq:44}
\end{equation}
where $\Id_3$ is the $3\times3$ identity matrix and $\tau\bm u^{\T}\bm u=\frac12\bm u^{\T}(2\tau\Id_3)\bm u$ is used. Thus, the weight only changes the quadratic term. Completing the square around a new center, the weighted distribution becomes a new Gaussian multiplied by a coefficient $L_{ij}$, which is produced by the completion of the square and does not depend on $\bm u$,
\begin{equation}
G_3(\bm u-\ub;\Sigma_{u,ij})\,e^{-\tau u^2}=L_{ij}\,G_3\bigl(\bm u-\ub^{(\tau)};\Sigma^{(\tau)}_{u,ij}\bigr),
\label{eq:47}
\end{equation}
where the new covariance and center follow from the quadratic and linear terms,
\begin{equation}
\bigl[\Sigma^{(\tau)}_{u,ij}\bigr]^{-1}=\Sigma_{u,ij}^{-1}+2\tau\Id_3,\qquad
\bigl[\Sigma^{(\tau)}_{u,ij}\bigr]^{-1}\ub^{(\tau)}=\Sigma_{u,ij}^{-1}\ub.
\label{eq:45}
\end{equation}
With $\Em_{ij}=\Id_3+2\tau\Sigma_{u,ij}$, Eq.~\eq{45} gives
\begin{equation}
\begin{aligned}
\Sigma^{(\tau)}_{u,ij}&=\Sigma_{u,ij}\Em_{ij}^{-1},\qquad
\ub^{(\tau)}=\Em_{ij}^{-1}\ub,\\
L_{ij}&=\frac{\exp\bigl(-\tau\,\ub^{\T}\Em_{ij}^{-1}\ub\bigr)}{\sqrt{\det\Em_{ij}}}.
\end{aligned}
\label{eq:46c}
\end{equation}
The two averages in Eq.~\eq{43} then follow directly from Eq.~\eq{47}.
Integrating Eq.~\eq{47} over $\bm u$, the new Gaussian integrates to unity, which gives the first average. Multiplying Eq.~\eq{47} by $u_z^2$ before the integration gives the second moment of the new Gaussian in the $z$ direction, i.e., its variance plus the squared mean. Thus,
\begin{equation}
\bigl\langle e^{-\tau u^2}\bigr\rangle=L_{ij},\qquad
\bigl\langle u_z^2e^{-\tau u^2}\bigr\rangle=L_{ij}\Bigl([\Sigma^{(\tau)}_{u,ij}]_{zz}+\bigl(\bar u^{(\tau)}_z\bigr)^2\Bigr).
\label{eq:48}
\end{equation}
By substituting Eq.~\eq{48} into Eq.~\eq{43}, the velocity factor can be expressed as
\begin{equation}
\Jzz^{ij}=\frac1{\sqrt\pi}\int_0^\infty\frac{\dd\tau}{\sqrt\tau}\,L_{ij}\Bigl\{1-6\tau\Bigl([\Sigma^{(\tau)}_{u,ij}]_{zz}+\bigl(\bar u^{(\tau)}_z\bigr)^2\Bigr)\Bigr\}.
\label{eq:49}
\end{equation}
Equation~\eq{49} reduces the three-dimensional velocity integral to a one-dimensional integral over $\tau$, while the finite longitudinal velocity spread, the nonzero mean relative velocity, and the $x$--$z$ correlation are retained. For $\ub=0$, Eq.~\eq{49} reduces to the B$\&$M formula~\cite{BM1983,KuboOide2001}.

Moreover, $\Delta$ enters only through $\ub=\bm b_{ij}\Delta$ in Eq.~\eq{49}, so that $\ub^{(\tau)}=\bm b^{(\tau)}_{ij}\Delta$ with $\bm b^{(\tau)}_{ij}=\Em_{ij}^{-1}\bm b_{ij}$. The exponential factor of $L_{ij}$ in Eq.~\eq{46c} and the factor $e^{-\alpha_{ij}\Delta^2/2}$ of $P_{ij}$ in Eq.~\eq{32} combine into $e^{-q_{ij}\Delta^2}$ with
\begin{equation}
q_{ij}(\tau;s)=\frac{\alpha_{ij}}2+\tau\,\bm b_{ij}^{\T}\Em_{ij}^{-1}\bm b_{ij}.
\label{eq:53}
\end{equation}
By substituting Eq.~\eq{49} into Eq.~\eq{41} and exchanging the order of the integrals, $A_{ij}$ can be expressed as
\begin{equation}
\begin{aligned}
A_{ij}(s)={}&\frac{C_{ij}}{\sqrt\pi}\int_0^\infty\frac{\dd\tau}{\sqrt\tau\sqrt{\det\Em_{ij}}}\int_{d_{ij}-L}^{d_{ij}+L}\omega_{ij}(\Delta)\,e^{-q_{ij}\Delta^2}\\
&\times\Bigl\{1-6\tau[\Sigma^{(\tau)}_{u,ij}]_{zz}-6\tau\bigl(b^{(\tau)}_{z,ij}\bigr)^2\Delta^2\Bigr\}\,\dd\Delta.
\end{aligned}
\label{eq:55}
\end{equation}
Thus, the average over $\Delta$ involves only the two integrals
\begin{equation}
\begin{aligned}
M_{0,ij}(q)&=\int\omega_{ij}(\Delta)\,e^{-q\Delta^2}\,\dd\Delta,\\
M_{2,ij}(q)&=\int\omega_{ij}(\Delta)\,\Delta^2e^{-q\Delta^2}\,\dd\Delta=-\frac{\dd M_{0,ij}}{\dd q}.
\end{aligned}
\label{eq:56}
\end{equation}
Since $\omega_{ij}$ is linear on the rising part $[d_{ij}-L,d_{ij}]$ and the falling part $[d_{ij},d_{ij}+L]$ of the triangle, both integrals can be expressed as
\begin{equation}
\begin{aligned}
M_{m,ij}(q)=\frac1{L^2}\Bigl\{&\bigl[(L-d_{ij})F_m+F_{m+1}\bigr]_{d_{ij}-L}^{d_{ij}}\\
&+\bigl[(L+d_{ij})F_m-F_{m+1}\bigr]_{d_{ij}}^{d_{ij}+L}\Bigr\},\qquad m=0,2,
\end{aligned}
\label{eq:59}
\end{equation}
where $[F]_a^b=F(b)-F(a)$, and the functions $F_n(\Delta;q)$ satisfy $\partial F_n/\partial\Delta=\Delta^ne^{-q\Delta^2}$,
\begin{equation}
\begin{aligned}
F_0&=\frac{\sqrt\pi}{2\sqrt q}\operatorname{erf}(\sqrt q\,\Delta), &F_1&=-\frac{e^{-q\Delta^2}}{2q},\\
F_2&=\frac{F_0}{2q}-\frac{\Delta e^{-q\Delta^2}}{2q}, &F_3&=-\frac{(1+q\Delta^2)e^{-q\Delta^2}}{2q^2}.
\end{aligned}
\label{eq:58}
\end{equation}
For $q\to0$, $M_{0,ij}\to1$ and $M_{2,ij}\to d_{ij}^2+L^2/6$.
Then, the $A_{ij}$ can be expressed as
\begin{equation}
\begin{aligned}
A_{ij}(s)={}&\frac{C_{ij}}{\sqrt\pi}\int_0^\infty\frac{\dd\tau}{\sqrt\tau\sqrt{\det\Em_{ij}}}\Bigl\{\Bigl[1-6\tau[\Sigma^{(\tau)}_{u,ij}]_{zz}\Bigr]M_{0,ij}(q_{ij})\\
&-6\tau\bigl(b^{(\tau)}_{z,ij}\bigr)^2M_{2,ij}(q_{ij})\Bigr\}.
\end{aligned}
\label{eq:62}
\end{equation}
Equation~\eq{62} carries out the integrals over the position, the two velocities, and $\Delta$ analytically, leaving only a one-dimensional numerical integral over $\tau$; the required $C_{ij}$, $\alpha_{ij}$, $\bm b_{ij}$, and $\Sigma_{u,ij}$ all follow from the current $\Snu$ of the two slices.

\section{Frozen Gaussian approximation}
\label{sec:FG}

\subsection{IBS growth from the slice model}
\label{sec:FGslice}

In this section, the frozen Gaussian approximation is considered for a bunch with a 6D Gaussian distribution at the entrance. In Eq.~\eq{2}, $S_{\delta,j}$ of each slice contains the scattering integrals with all slices, so that $N^2$ scattering integrals $A_{ij}$ have to be evaluated at each position, where $N$ denotes the number of slices. In the frozen Gaussian approximation, the bunch is kept in the form of a 6D Gaussian distribution, and the mismatch between the slices is neglected. Then, the scattering integral between slice $j$ and the whole bunch, $\bar A_j=\sum_iW_iA_{ij}$, can be evaluated analytically, where $W_i$ is the fraction of particles in slice $i$. Moreover, in the frozen case, the IBS growth of each slice is proportional to the local current, so that only the case of $z=0$ has to be evaluated, and the result is scaled to any $z$. For a Gaussian bunch, the longitudinal density at the entrance is
\begin{equation}
\lambda_0(z_0)=\frac{1}{\sqrt{2\pi}\,\sigma_{z0}}\exp\Bigl(-\frac{z_0^2}{2\sigma_{z0}^2}\Bigr),
\label{eq:FG1}
\end{equation}
where $\sigma_{z0}$ is the rms bunch length. For a 6D Gaussian bunch, the particles with a given $z_0$ also follow a Gaussian distribution, whose centroid $\bm d(0)z_0$ depends on $z_0$, whereas the second-moment matrix $\Snu$ around this centroid does not. Thus, all slices have the same $\Snu$ at the entrance, and without IBS they remain equal under the linear transport in Eq.~\eq{19}.

Since IBS leads to different energy-spread growth in different slices, $\Snu$ differs between the slices at any $s$, and in the frozen Gaussian approximation, the second-moment matrix of the bunch seen by the collisions is taken as the weighted average $\overline\Sigma_\nu=\sum_jW_j\Snu$. With the density $\mathcal{I}_j\lambda_0/W_j$ of $z_0$ within each slice instead of the uniform $\mathcal{I}_j/L$ in Eq.~\eq{15}, the factor $1/L^2$ in Eq.~\eq{5} becomes $\lambda_0(z_{01})\lambda_0(z_{02})/(W_iW_j)$. By summing over all slices $i$ with the weights $W_i$, the distribution of particle 1 can be expressed as $\sum_iW_if_i=G_6(\bm X;\Sigma_X)$, where $\Sigma_X=\overline\Sigma_\nu+\sigma_{z0}^2\bm d\bm d^{\T}$ is the second-moment matrix of the whole bunch. As in Eq.~\eq{22}, the first term of $\Sigma_X$ is the second-moment matrix for a given $z_{01}$, and the second term arises from the distribution of the centroid $\bm dz_{01}$, with $L^2/12$ replaced by $\sigma_{z0}^2$. Since the particles of slice $j$ now collide with the whole bunch instead of a single slice $i$ in this case, the formulas of Sec.~\ref{sec:II} apply with $\Sigma_{\nu,i}$ replaced by $\Sigma_X$ of the whole bunch, $\Sigma_{\nu,j}$ replaced by $\overline\Sigma_\nu$, and $\Delta$ replaced by $-z_{02}$ ($\langle z_{01}\rangle=0$ for the whole bunch). That is, the sum of the second-moment matrices of the two slices in Eqs.~\eq{Sigrij} and \eq{27} becomes $\Sigma_X+\overline\Sigma_\nu$, and Eqs.~\eq{alphaC}, \eq{S9}, and \eq{33} give
\begin{equation}
\begin{aligned}
C_B&=\frac{\bar C}{\sqrt{1+\sigma_{z0}^2\bar\alpha}},\qquad
\alpha_B=\frac{\bar\alpha}{1+\sigma_{z0}^2\bar\alpha},\\
\bm b_B&=\frac{\bar{\bm b}}{1+\sigma_{z0}^2\bar\alpha},\qquad
\Sigma_{u,B}=\bar\Sigma_u+\frac{\sigma_{z0}^2\,\bar{\bm b}\bar{\bm b}^{\T}}{1+\sigma_{z0}^2\bar\alpha}.
\end{aligned}
\label{eq:FG2}
\end{equation}
Here, $\bar C$, $\bar\alpha$, $\bar{\bm b}$, and $\bar\Sigma_u$ are the values of Eqs.~\eq{alphaC} and \eq{33} for $\Sigma_{\nu,i}=\Sigma_{\nu,j}=\overline\Sigma_\nu$, i.e., for two particles from the same average slice, and the subscript $B$ indicates the collisions with the whole bunch. Notice that the whole bunch includes slice $j$ itself, and these parameters are the same for any slice $j$, so that $\bar A_j$ depends on $j$ only through the average of the position overlap factor and the velocity factor over $z_{02}$ within $I_j$. Since the average over $z_{01}$ is already contained in $\Sigma_X$, $\omega_{ij}$ in Eq.~\eq{41} is replaced by $\lambda_0(z_{02})/W_j$ of $z_{02}$ within slice $j$, and $\bar A_j$ can be expressed as
\begin{equation}
\bar A_j(s)=\frac{C_B}{W_j}\int_{I_j}\lambda_0(z_{02})\,e^{-\alpha_Bz_{02}^2/2}\,\Jzz^{B}(z_{02};s)\,\dd z_{02},
\label{eq:FG3}
\end{equation}
As in Eq.~\eq{55}, by substituting Eq.~\eq{49} into Eq.~\eq{FG3}, the average over $z_{02}$ involves only two integrals, but with the Gaussian density $\lambda_0(z_{02})/W_j$ within slice $j$ instead of the triangular $\omega_{ij}(\Delta)$, so that $M_{0,ij}$ and $M_{2,ij}$ in Eq.~\eq{56} are modified to
\begin{equation}
\begin{aligned}
M_{0,j}(q)&=\frac{\bigl[\operatorname{erf}(\sqrt{\tilde q}\,z_{02})\bigr]_{a_j}^{b_j}}{2W_j\sqrt{1+2\sigma_{z0}^2q}},\\
M_{2,j}(q)&=\frac{M_{0,j}}{2\tilde q}-\frac{\bigl[z_{02}e^{-\tilde qz_{02}^2}\bigr]_{a_j}^{b_j}}{2\sqrt{2\pi}\,\tilde q\,\sigma_{z0}W_j},
\end{aligned}
\label{eq:FG4}
\end{equation}
where $\tilde q=q+1/(2\sigma_{z0}^2)$, $a_j$ and $b_j$ are the edges of $I_j$, and the integration bounds are extended to $\pm\infty$. Thus, by replacing $M_{0,ij}$ and $M_{2,ij}$ in Eq.~\eq{62} with $M_{0,j}$ and $M_{2,j}$ and the subscript $ij$ with $B$, $\bar A_j$ can be expressed as
\begin{equation}
\begin{aligned}
\bar A_j(s)={}&\frac{C_B}{\sqrt\pi}\int_0^\infty\frac{\dd\tau}{\sqrt\tau\sqrt{\det\Em_B}}\Bigl\{\Bigl[1-6\tau[\Sigma^{(\tau)}_{u,B}]_{zz}\Bigr]M_{0,j}(q_B)\\
&-6\tau\bigl(b^{(\tau)}_{z,B}\bigr)^2M_{2,j}(q_B)\Bigr\},
\end{aligned}
\label{eq:FG5}
\end{equation}
where $\Em_B$, $\Sigma^{(\tau)}_{u,B}$, $\bm b^{(\tau)}_B=\Em_B^{-1}\bm b_B$, and $q_B$ follow from Eqs.~\eq{46c} and \eq{53} with the subscript $ij$ replaced by $B$.

\subsection{IBS growth from the B\&M model}
\label{sec:FGBM}
Since the bunch is kept in the form of a 6D Gaussian distribution in the frozen Gaussian approximation, the IBS growth of the whole bunch can also be obtained from the B\&M model for a Gaussian bunch~\cite{BM1983,Bane2002}, and it is derived in the following to be compared with that from the slice model.
In the B\&M model, the momentum distribution of the bunch is described by the matrix $L$ and the coefficient $A$,
\begin{equation}
\begin{aligned}
L&=\begin{pmatrix}
[\Sigma_X^{-1}]_{22}&0&\gamma_0[\Sigma_X^{-1}]_{26}\\
0&[\Sigma_X^{-1}]_{44}&0\\
\gamma_0[\Sigma_X^{-1}]_{26}&0&\gamma_0^2[\Sigma_X^{-1}]_{66}
\end{pmatrix},\\
A&=\frac{r_e^2cN_e}{64\pi^2\beta_0^3\gamma_0^4\sqrt{\det\Sigma_X}},
\end{aligned}
\label{eq:BM1}
\end{equation}
where $[\Sigma_X^{-1}]_{ij}$ is the momentum block of $\Sigma_X^{-1}$, and $N_e$ is the number of particles in the bunch; $L$ is the matrix in Refs.~\cite{Bane2002,Pan2025} written for a general Gaussian bunch, and $\sqrt{\det\Sigma_X}$ reduces to $\epsilon_x\epsilon_y\sigma_z\sigma_\delta$ for an uncoupled bunch without dispersion and chirp. The matrix integral in the B\&M model is
\begin{equation}
B=\int_0^\infty\frac{\sqrt\lambda\,\dd\lambda}{\sqrt{\det(L+\lambda\Id_3)}}\Bigl\{\operatorname{Tr}\bigl[(L+\lambda\Id_3)^{-1}\bigr]\Id_3-3(L+\lambda\Id_3)^{-1}\Bigr\}.
\label{eq:BM2}
\end{equation}
For an uncoupled bunch without chirp, the B\&M model gives the growth rate of the momentum spread in a storage ring, $1/T_p=\sigma_\delta^{-1}\dd\sigma_\delta/\dd t=4\pi A\ln\Lambda\,\gamma_0^2[B]_{33}/\sigma_\delta^2$, where the synchrotron oscillation transfers half of the momentum diffusion to $z$; thus, the diffusion rate of $\langle\delta^2\rangle$ caused by the collisions is $4\sigma_\delta^2/T_p$, where one factor of 2 comes from the conversion from $\sigma_\delta$ to $\sigma_\delta^2$ and the other from the synchrotron oscillation. In a single pass, there is no synchrotron oscillation, so that this diffusion remains in $\delta$, and its rate per unit length is $4\sigma_\delta^2/(\beta_0cT_p)$; since $\sigma_\delta^2$ cancels with that in $1/T_p$, the rate depends only on $L$ and $A$, and for a Gaussian bunch, the B\&M model gives
\begin{equation}
S_\delta^{\mathrm{B\&M}}=\frac{16\pi A\ln\Lambda\,\gamma_0^2}{\beta_0c}\,[B]_{33}.
\label{eq:BM3}
\end{equation}
On the other hand, the sum of Eq.~\eq{FG5} over all slices with the weights $W_j$ removes the slice label of particle 2, so that both particles belong to the whole bunch $G_6(\bm X;\Sigma_X)$, and
\begin{equation}
\begin{aligned}
\sum_jW_j\bar A_j&=C_X\,\Jzz(0;\Sigma_{u,X}),\qquad
C_X=\frac{1}{(2\pi)^{3/2}\sqrt{\det(2\Sigma_{X,r})}},\\
\Sigma_{u,X}&=2T_v\bigl(\Sigma_{X,P}-\Sigma_{X,Pr}\Sigma_{X,r}^{-1}\Sigma_{X,Pr}^{\T}\bigr)T_v^{\T}.
\end{aligned}
\label{eq:FG6}
\end{equation}
where $C_X$ and $\Sigma_{u,X}$ are the values of Eqs.~\eq{alphaC} and \eq{33} for $\Sigma_{\nu,i}=\Sigma_{\nu,j}=\Sigma_X$. For a single Gaussian distribution, the momentum distribution at any $s$ has a second-moment matrix independent of the position and a mean linear in the position, so that two particles at the same position have zero mean relative velocity and the relative-velocity second-moment matrix $\Sigma_{u,X}$ everywhere. Since $\Sigma_{u,X}=2\beta_0^2\gamma_0^2c^2L^{-1}$, by setting $\ub=0$ in Eq.~\eq{49} and substituting $\lambda=2\tau$, Eqs.~\eq{FG6} and \eq{2} give
\begin{equation}
\sum_jW_jS_{\delta,j}=\frac{4\pi r_e^2c\ln\Lambda\,N_e}{\beta_0^3\gamma_0^2}\,C_X\,\Jzz(0;\Sigma_{u,X})=S_\delta^{\mathrm{B\&M}}.
\label{eq:BM4}
\end{equation}
Thus, the bunch average of the frozen Gaussian approximation is identical to the B\&M model for the whole bunch, and Eq.~\eq{FG5} distributes this growth among the slices. 

In this section, the effect of IBS on the energy spread in the frozen Gaussian approximation is derived based on the slice model in this study and on the B\&M model, respectively, and the comparison of the two is given in the following sections.

\section{Simulation for the bunch with energy chirp}
\label{sec:sim}

In the previous sections, the theoretical model has been introduced for both the Gaussian bunch and more realistic cases. In this section, the design of the simulations is considered. For IBS in linacs, including single-pass and recirculating linacs, the commonly used methods besides the analytical calculation are \textsc{elegant}~\cite{Borland2000,XiaoBorland2009} and PIC codes such as RF-Track~\cite{DesireValdor2024,Desire2025}. However, for a bunch with an energy chirp passing through a beamline with nonzero dispersion, large differences appear between the results of the two codes, which is also the motivation of this work.

\begin{figure}
    \centering
    \includegraphics[width=1\linewidth]{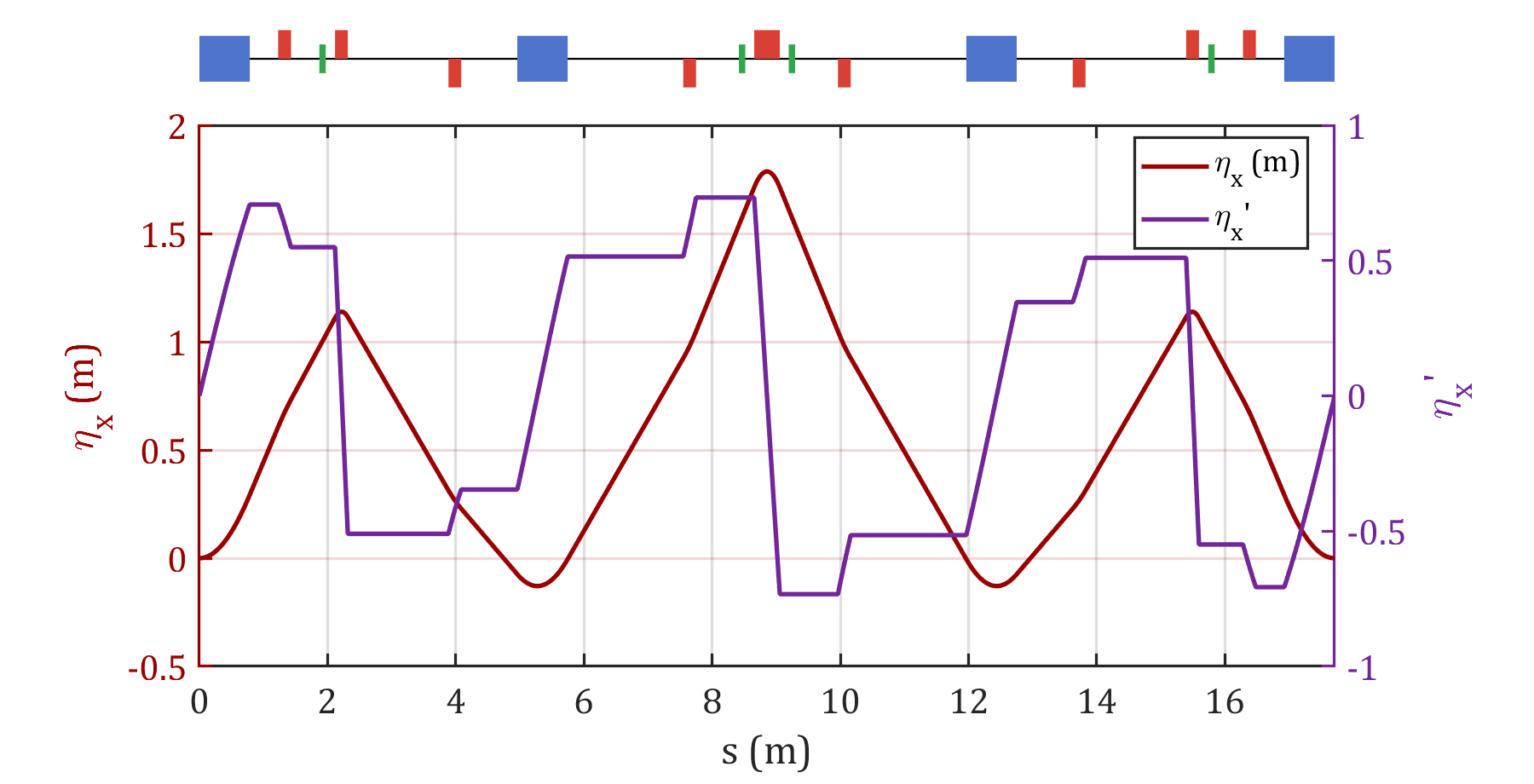}\\
    \includegraphics[width=1\linewidth]{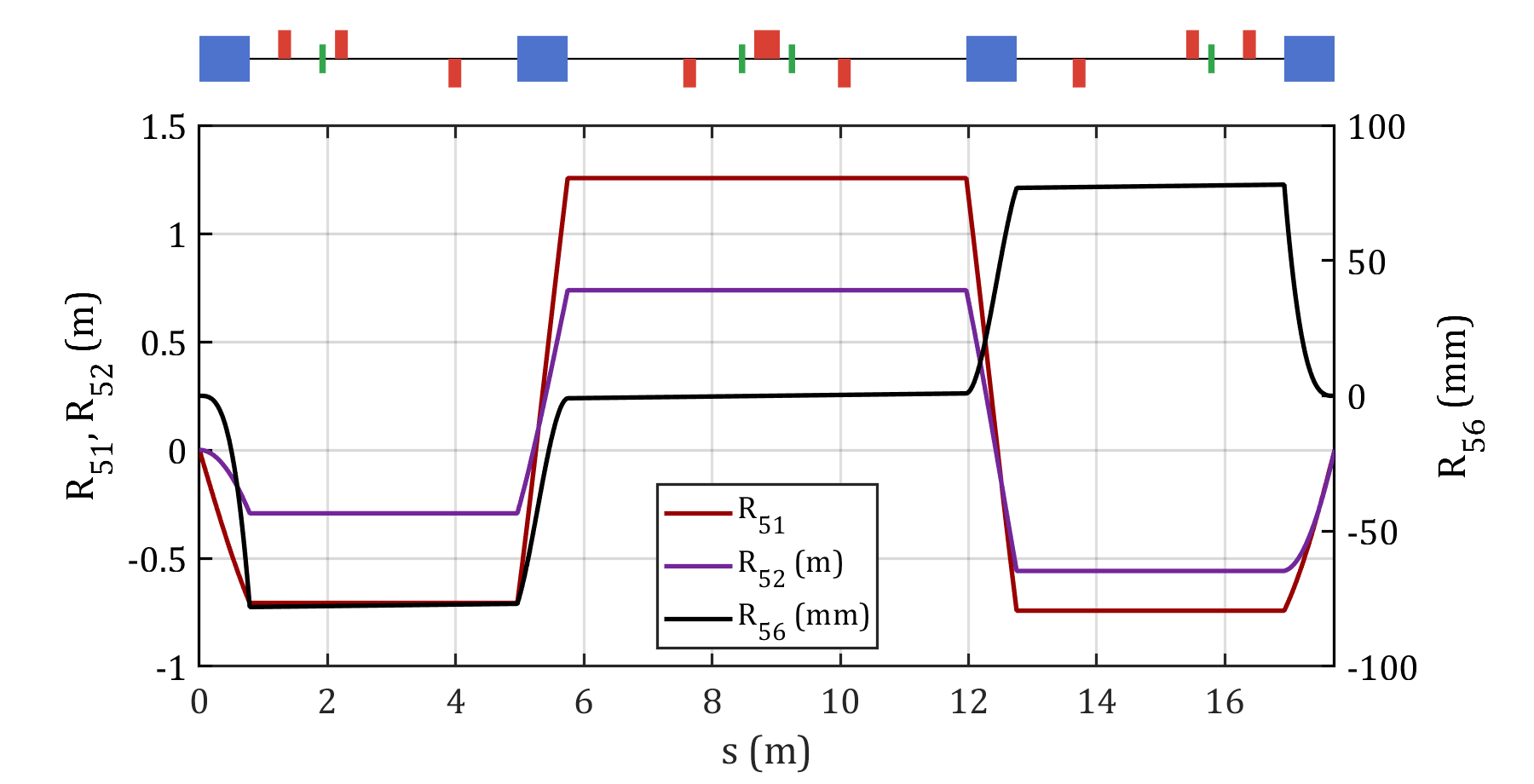}
    \caption{Optical parameters of the arc. (a) Transfer-matrix elements $R_{51}$, $R_{52}$, and $R_{56}$ from the entrance to the position $s$. (b) Horizontal dispersion $\eta_x$ and its derivative $\eta_x'$. The layout of the arc is shown at the top of each panel: dipoles (blue), quadrupoles (red; above and below the axis for horizontal focusing and defocusing, respectively), and sextupoles (green).}
    \label{fig:arcoptics}
\end{figure}

Some of the optical parameters of the arc are shown in Fig.~\ref{fig:arcoptics}. The reference energy is 30~MeV, so that the longitudinal compression caused by the drift sections through the velocity slippage cannot be neglected. The arc consists of four identical dipoles with a total deflecting angle of 180$^\circ$, and quadrupoles are placed between the dipoles to make the arc achromatic. Moreover, the total $R_{56}$ of the arc, including the velocity slippage, is set to zero. This is because large initial chirps are used in the tests, which would lead to a strong compression for a residual $R_{56}$; to observe the change of the UES caused by IBS, the growth of the UES due to the compression has to be avoided. For the same reason, sextupoles are included to make $T_{566}$ of the arc close to zero, so that the change of the UES caused by the nonlinear compression, which varies along the bunch, is also avoided. 

In \textsc{elegant}, the bunch is divided into slices along the current longitudinal position, the IBS growth rate of each slice is computed with the B\&M formula from the slice parameters~\cite{Borland2000,XiaoBorland2009}. This implementation has been compared with a semi-analytical IBS calculation in a recirculating beamline, and consistent growth rates were obtained~\cite{Tsai2020}. RF-Track uses a Monte Carlo method, in which each macroparticle collides with an average particle sampled from the local density, mean velocity, and velocity spread stored on a 3D mesh, and the scattering angle is sampled from the Rutherford cross section~\cite{DesireValdor2024}. RF-Track has been benchmarked against REPTIL for the European XFEL injector~\cite{Gjonaj2022} and against a semi-analytical model~\cite{DiMitri2020,PerosaDiMitri2021}, and good agreement in the UES growth was found~\cite{Desire2025}. As discussed in Sec.\ref{sec:intro}, the previous benchmarks of IBS in \textsc{elegant} and RF-Track were performed in drift sections or for bunches without an energy chirp, and the case of a strongly chirped bunch in a dispersive arc has not been examined.

\begin{table}[htbp]
\caption{Parameters of the Gaussian bunch at the entrance of the arc.}
\label{tab:bunch}
\begin{ruledtabular}
\begin{tabular}{lcc}
Parameter & Value & Unit \\
\hline
Reference energy & 30 & MeV \\
Bunch charge & 100 & pC \\
Rms bunch length $\sigma_{z0}$ & 0.3 & mm \\
Normalized emittance $\epsilon_{n,x}$, $\epsilon_{n,y}$ & 0.2, 0.2 & $\mu$m \\
Initial UES &  2 & keV \\
Linear energy chirp $h$ & 0--20 & m$^{-1}$ \\
$\beta_x$, $\alpha_x$ & 1.259, 1.370 & m, -- \\
$\beta_y$, $\alpha_y$ & 1.555, 1.609 & m, -- \\
\end{tabular}
\end{ruledtabular}
\end{table}

Since few references are available for such simulations, convergence tests of the simulations are necessary. The tests use the ideal Gaussian bunch listed in Table~\ref{tab:bunch} with $10^5$ macroparticles, and the lattice is the arc discussed above.

\begin{figure}
    \centering
    \includegraphics[width=1\linewidth]{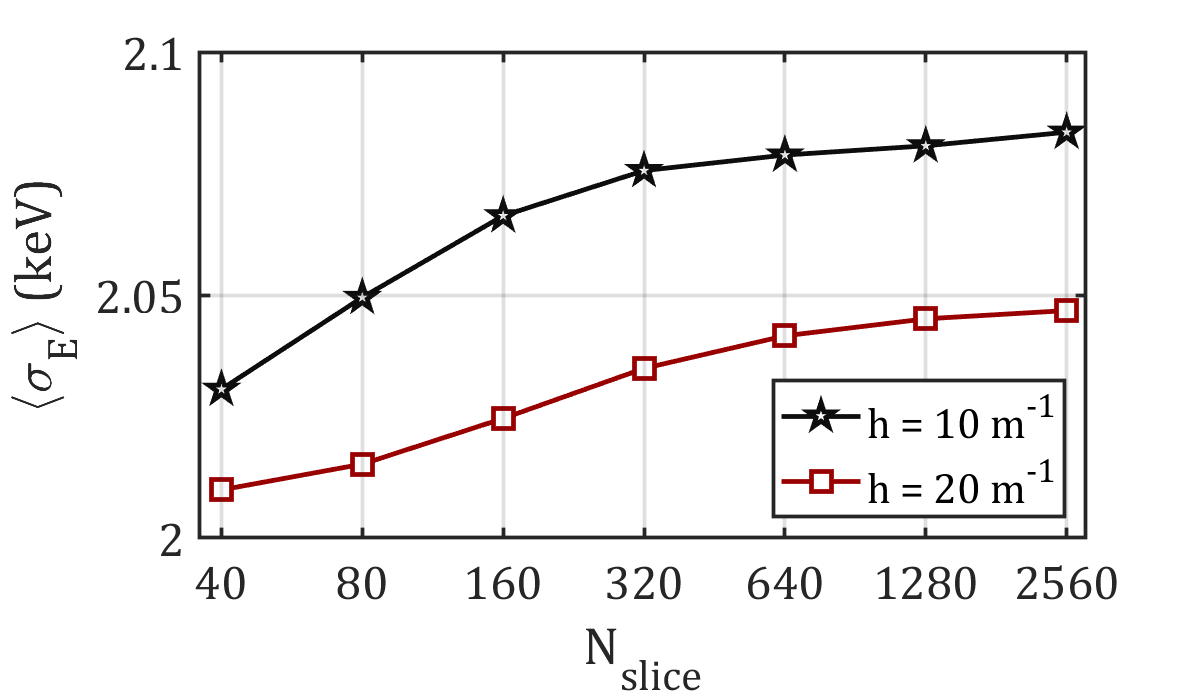}
    \caption{Weighted UES at the exit of the arc given by \textsc{elegant} as a function of the number of slices $N_{\mathrm{slice}}$ used for the IBS calculation, for $h=10$ and $20$~m$^{-1}$.}
    \label{fig:elescan}
\end{figure}

For \textsc{elegant}, a convergence test on the number of longitudinal slices is performed~\cite{DiMitri2014}, so that the correlated energy spread of the chirped bunch within each slice is not counted in the UES, which would otherwise overestimate the UES and affect the IBS calculation.  As shown in Fig.~\ref{fig:elescan}, the weighted UES at the exit given by \textsc{elegant} converges as the number of slices increases, and the change is below 0.2\% for each doubling beyond 640 slices at both $h=10$ and $20$~m$^{-1}$. However, the growth of the UES due to IBS after the convergence is negligible: the weighted UES increases only from 2.00~keV without IBS to 2.08~keV at $h=10$~m$^{-1}$ and to 2.05~keV at $h=20$~m$^{-1}$. Then, to avoid the influence of the numerical noise, the number of slices in \textsc{elegant} is set to 640 in the following simulations.

\begin{figure}
    \centering
    \includegraphics[width=1\linewidth]{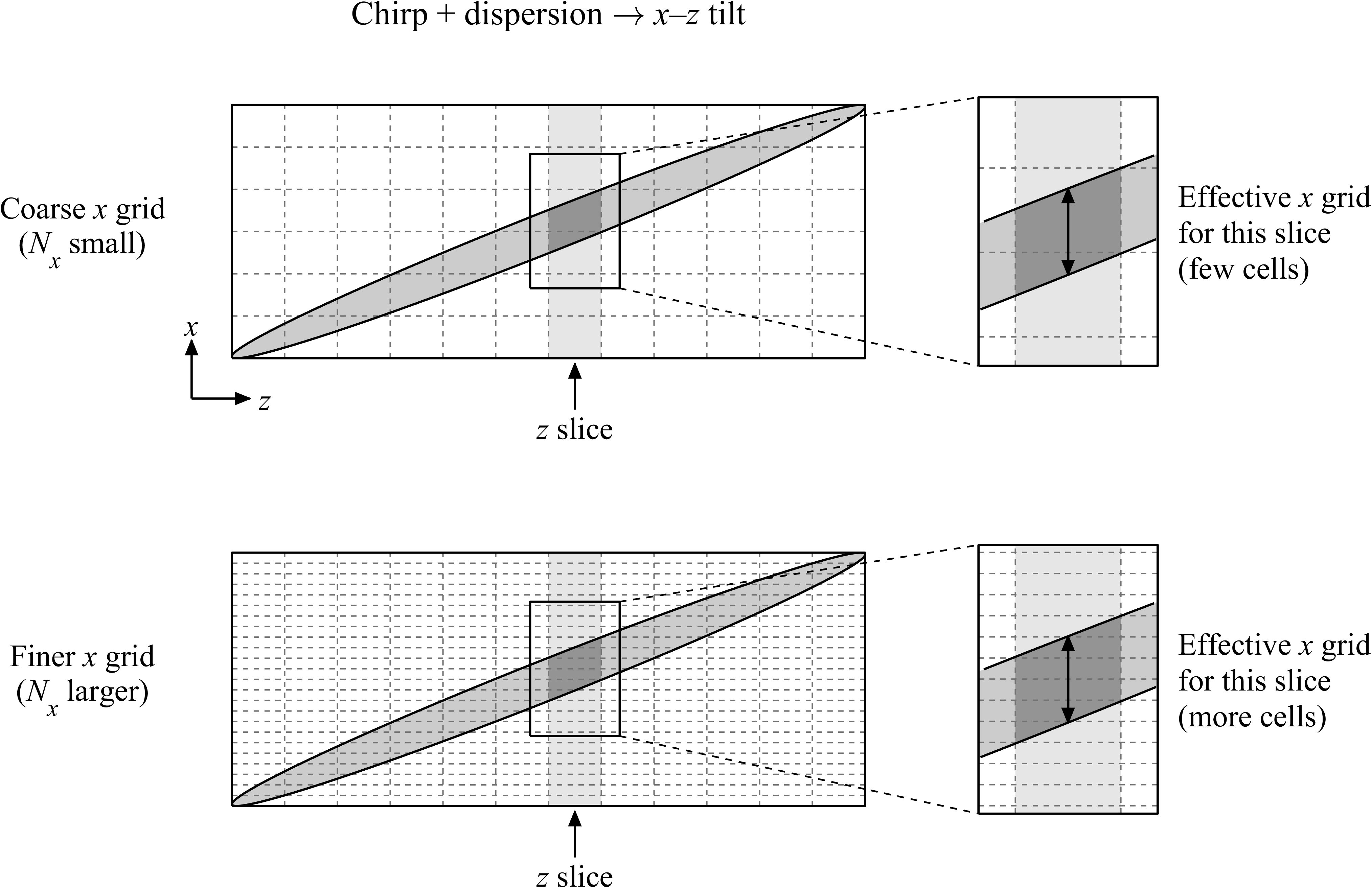}
    \caption{Schematic of the 3D mesh for a bunch with an $x$--$z$ tilt caused by the chirp and the dispersion. The mesh is generated according to the projected size of the bunch, so that a longitudinal slice occupies only a few transverse cells for a coarse $x$ mesh (top), and more cells for a finer $x$ mesh (bottom)}
    \label{fig:mesh}
\end{figure}

\begin{figure}
    \centering
    \includegraphics[width=1\linewidth]{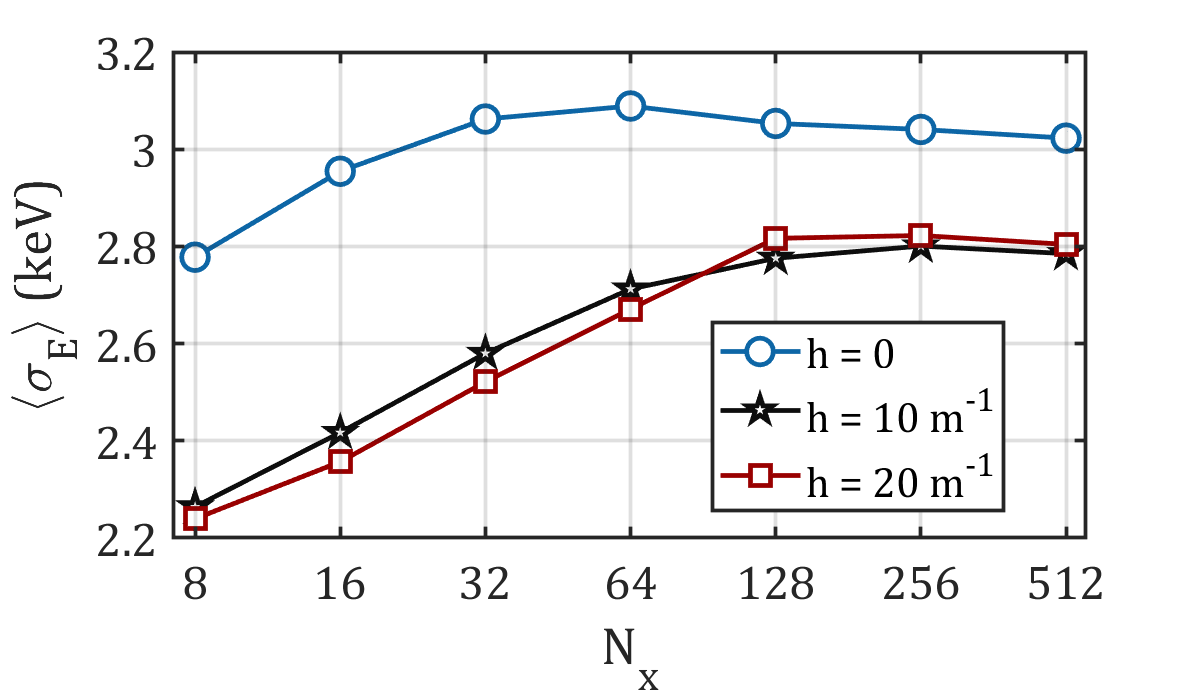}
    \caption{Weighted UES at the exit of the arc given by RF-Track as a function of the number of transverse mesh points $N_x$, for $h=0$, $10$, and $20$~m$^{-1}$, with $N_y=32$ and $N_z=64$.}
    \label{fig:rftscan}
\end{figure}

RF-Track is a 3D PIC code with a 3D mesh. The correlated energy spread along $z$ appears to be removed, so that the results converge well with the number of longitudinal mesh points (not shown), and $N_z=64$ is used, which is a common setting in 3D PIC simulations. Moreover, for a bunch with a large chirp, an $x$--$z$ tilt is generated where the dispersion is nonzero. As shown in Fig.~\ref{fig:mesh}, the 3D mesh is generated only according to the projected size of the bunch, so that a longitudinal slice occupies only a few cells in the transverse direction, which leads to a large error due to the reduced number of sampling points. To find the optimal setting, the results with different $N_x$ are compared for $h=0$, $10$, and $20$~m$^{-1}$ in Fig.~\ref{fig:rftscan}. The weighted UES converges as $N_x$ increases: it converges from $N_x=128$ for $h=10$ and $20$~m$^{-1}$, whereas it already converges at $N_x=32$ for $h=0$. That is, the convergence is easier for $h=0$ than for a large chirp, since the $x$--$z$ tilt caused by the dispersion lets a longitudinal slice occupy only a few transverse cells for a large chirp. Thus, $N_x=256$ is used in the following simulations.

Notice that, for the same lattice and bunch, \textsc{elegant} and RF-Track give very different results for a large chirp after the convergence tests. For \textsc{elegant}, the growth of the UES due to IBS is almost negligible, whereas for RF-Track, the growth is not negligible; e.g., at $h=10$~m$^{-1}$, the weighted UES at the exit is 2.08~keV for \textsc{elegant} and 2.80~keV for RF-Track, compared with 2.00~keV without IBS. In the following sections, a more detailed benchmark of the theoretical predictions, RF-Track, and \textsc{elegant} is presented for both the ideal Gaussian bunch and more realistic bunches.

\section{Comparison between the theory and simulations}
\label{sec:7}
In the previous sections, the slice model and the simulation settings have been introduced. In this section, the slice model is benchmarked against elegant and RF-Track. Both the Gaussian bunch and a more realistic bunch are considered. The focus is on the evolution of the bunch in a beamline with dispersion. A conventional straight section without dispersion is not considered here. In such a beamline, there is no transverse--longitudinal coupling, and the calculation is straightforward. The corresponding discussion is given in Appendix~\ref{app:straight}. 

In the slice model, the IBS growth in Eq.~\eq{2} is applied as kicks along the beamline. The beamline is divided into steps of length $\Delta s$, and in each step, $\Snu$ is first transported over $\Delta s/2$ with $R(s)$ in Eq.~\eq{19}, then each slice receives the kick
\begin{equation}
[\Snu]_{66}\;\to\;[\Snu]_{66}+S_{\delta,j}\,\Delta s,
\label{eq:kick}
\end{equation}
and finally $\Snu$ is transported over the remaining $\Delta s/2$. Since $S_{\delta,j}$ at each kick is evaluated with the current $\Snu$, which contains all previous kicks, the UES generated by IBS feeds back into the subsequent growth, and the kicks in $\delta$ are also transferred to $z$ and $x$ through $R_{56}$ and $R_{16}$ in the subsequent transport, in the same way as the energy kicks of coherent synchrotron radiation in a beamline with dispersion~\cite{Chen2026CSR}.

This section consists of two parts. In the first part, IBS in the ideal Gaussian bunch is considered, and the two methods in Sec.~\ref{sec:FG} are compared with the simulations. In the second part, a more realistic bunch is considered, and the slice model in Sec.~\ref{sec:III} is compared with the simulations.

\subsection{IBS for the frozen Gaussian bunch}
\label{sec:cmpFG}
In the frozen Gaussian approximation, $\sum_iq_iA_{ij}$ in Eq.~\eq{2} is replaced by $eN_e\bar A_j$ with $\bar A_j$ in Eq.~\eq{FG5}, and $\overline\Sigma_\nu$ and $\Sigma_X$ are evaluated with the kicked $\Snu$ at each step. In Sec.~\ref{sec:sim}, the convergence of \textsc{elegant} and RF-Track has been tested. After the convergence, the growths of the UES caused by IBS given by the two codes still differ significantly.

\begin{figure}
    \centering
    \includegraphics[width=1\linewidth]{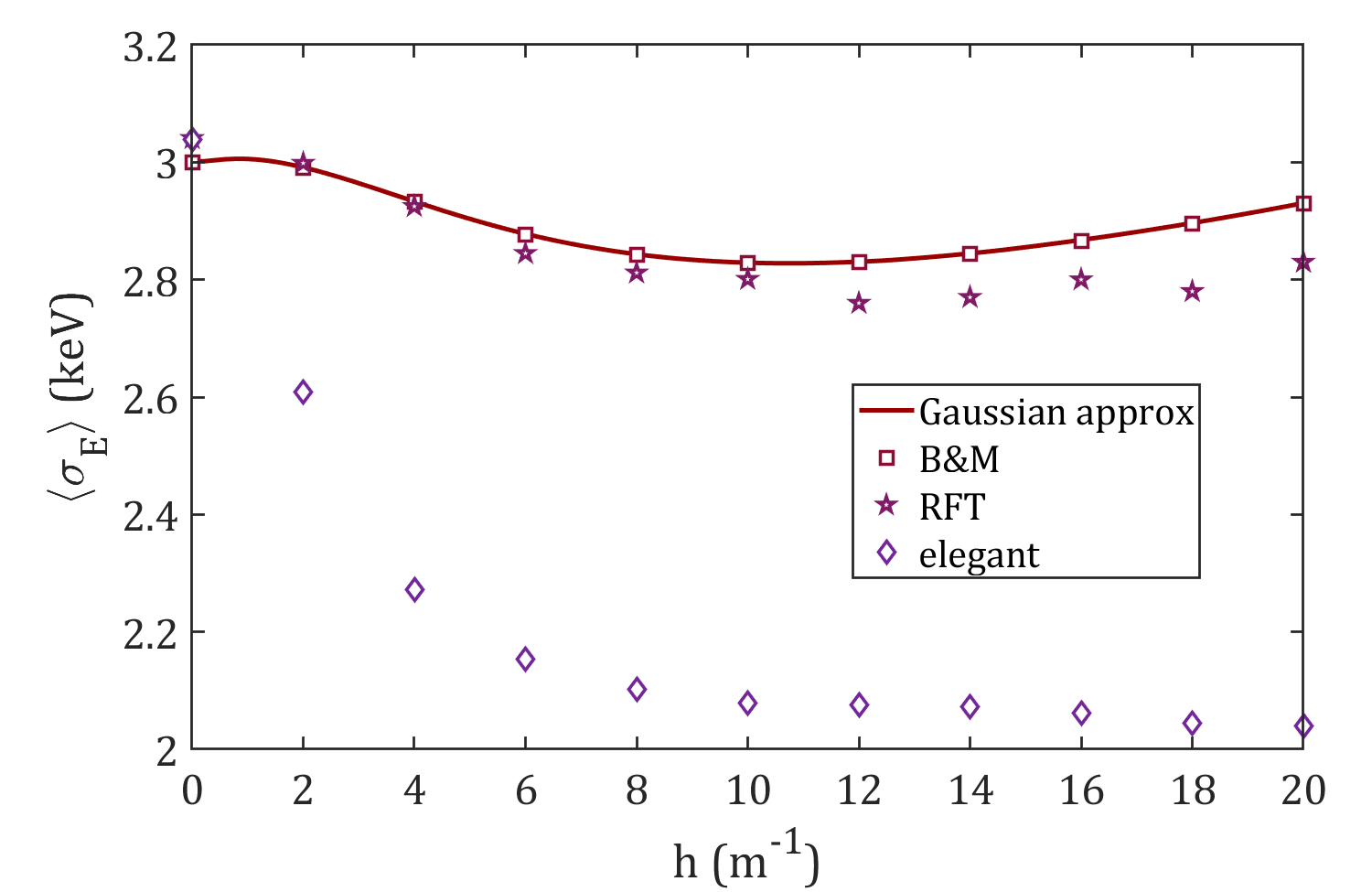}
    \caption{Weighted UES at the exit of the arc as a function of the chirp $h$ for the Gaussian bunch in Table~\ref{tab:bunch}. The curve labeled $\bar A_j$ is given by the slice model in the frozen Gaussian approximation, Eq.~\eq{FG5}, and the markers labeled $S_\delta^{\mathrm{B\&M}}$ are given by the B\&M model, Eq.~\eq{BM3}. The other markers are given by RF-Track and \textsc{elegant} with the settings in Sec.~\ref{sec:sim}.}
    \label{fig:gaubench}
\end{figure}

In this section, with the same beamline and the beam parameters in Table~\ref{tab:bunch}, a more detailed comparison is shown in Fig.~\ref{fig:gaubench}. In the detailed simulations, \textsc{elegant} shows an obvious difference from the other methods. Although all methods give close UES at $h=0$, a large difference remains once the chirp is introduced. This difference may come from the energy spread evaluated within each slice in \textsc{elegant}, where the slices are cut along the current $z$. As discussed in Sec.~\ref{sec:II}, the particles slip between such slices because of $R_{5i}$, which increases the energy spread within a single slice. Increasing the number of slices only reduces the correlation between the energy and $z$, whereas the correlation between the energy and the transverse coordinates remains. Thus, the UES is overestimated, and the effect of IBS is underestimated. 

The other three methods agree well over the whole range of $h$ or a part of it. The two theoretical methods give almost identical results. This is because the two methods are based on the same physical model and differ only in the mathematical expressions. For $h\le10$~m$^{-1}$, the two theoretical methods agree well with RF-Track. As $h$ increases further, the UES given by RF-Track remains nearly constant, whereas the theoretical prediction increases again. The deviation may come from two aspects. From the theoretical side, only the linear chirp and the linear transfer matrix are considered. Although the over-compression does not cause any error in the derivation, the bunch reaches the full compression inside the arc at $h\approx12.8$~m$^{-1}$ and becomes extremely short locally, which may lead to a large local IBS growth. From the simulation side, the higher-order effects are included in RF-Track, so that the bunch is not compressed to such an ultrashort length. Moreover, the $x$--$z$ tilt is large where the bunch is short, so that the mesh settings in Sec.~\ref{sec:sim} may not be sufficient there.

\subsection{IBS for the non-ideal bunch}

\begin{figure}
    \centering
    \includegraphics[width=1\linewidth]{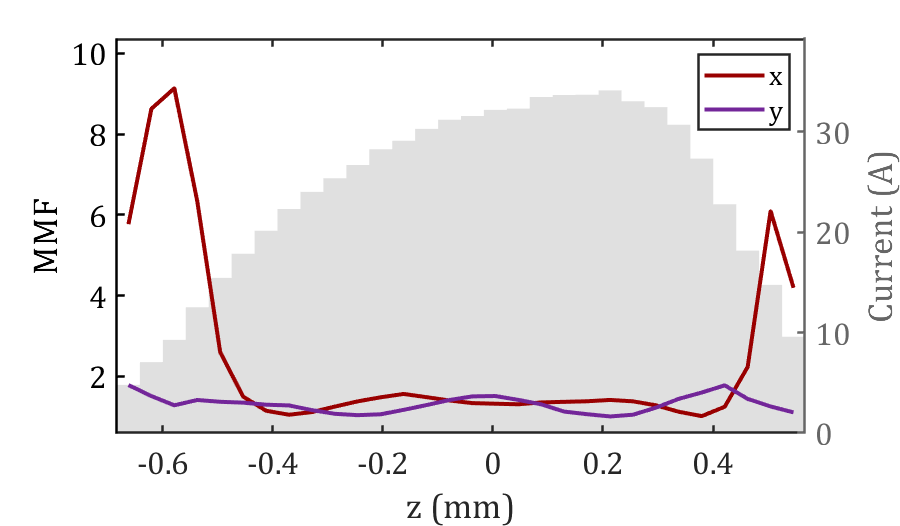}
    \caption{Mismatch factor (MMF) of each slice at the entrance of the arc for the realistic bunch in the horizontal ($x$) and vertical ($y$) planes, with respect to the projected Twiss parameters; $\mathrm{MMF}=1$ corresponds to a matched slice.}
    \label{fig:actbeam}
\end{figure}

For a more realistic bunch, the transverse and longitudinal distributions are taken from the Ref.~\cite{Chen2025NIMA}. In the transverse planes, the mismatch between the slices of the original bunch is preserved, and the distribution is not modified into a Gaussian distribution. In the longitudinal phase space, the original longitudinal profile is kept, the energy chirp is set as a linear chirp adjusted according to the actual simulation, and the initial UES is about 2~keV. The bunch charge, the rms bunch length, and the initial Twiss functions and other parameters are the same as those of the Gaussian bunch in Table~\ref{tab:bunch}, while the projected normalized emittances are 0.37 and 0.31~$\mu$m in $x$ and $y$, respectively. The mismatch factor (MMF) of each slice, $\mathrm{MMF}=(\beta\gamma_r-2\alpha\alpha_r+\gamma\beta_r)/2$, where the subscript $r$ denotes the projected Twiss parameters, and the longitudinal profile are shown in Fig.~\ref{fig:actbeam}.

\begin{figure}
    \centering
    \includegraphics[width=1\linewidth]{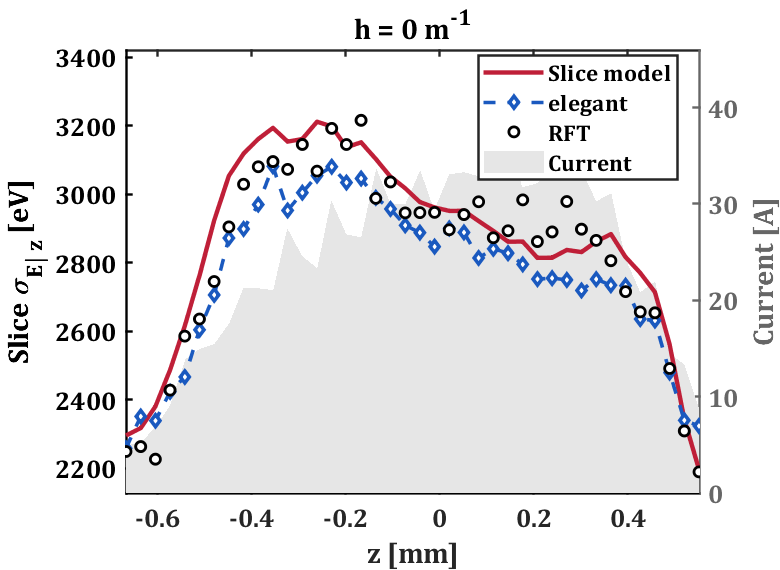}\\
    \includegraphics[width=1\linewidth]{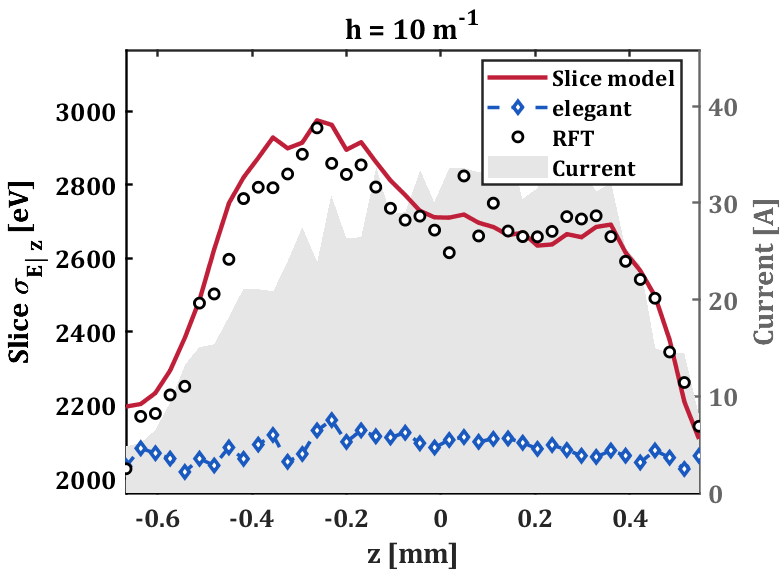}
    \caption{UES along the bunch at the exit of the arc for the realistic bunch, with $h=0$ (top) and $h=10$~m$^{-1}$ (bottom), given by the slice model based on Eq.~\eq{62}, RF-Track, and \textsc{elegant}.}
    \label{fig:actbench}
\end{figure}

As shown in Fig.~\ref{fig:actbeam}, in the core of the bunch, where the current is above half of the peak current, the MMF is about 1.0--2.2 in $x$, whereas it increases up to about 9 at the head and tail. In the $y$ plane, the MMF is between 1.0 and 1.8 along the whole bunch. The benchmark for the realistic bunch also includes the slice model, RF-Track, and \textsc{elegant}, whereas the B\&M formula is not included, since it applies only to a Gaussian bunch. The lattice parameters are the same as those in the previous subsection. For the realistic bunch, the distribution of the UES along the longitudinal slices is of more interest than the weighted UES of the whole bunch, especially the UES of the slices around peak-current. In this case, the IBS of each slice is not determined only by its local current, as in the 6D Gaussian case, but also by the envelope evolution of each slice and the local collision integrals between mismatched slices; thus, $A_{ij}$ has to be calculated separately for each slice pair, as in Eq.~\eq{62}. 

The benchmarks for $h=0$ and $10$~m$^{-1}$ are shown in Fig.~\ref{fig:actbench}. For $h=0$ (top), the three methods give similar results, and the slice model is closer to RF-Track. The growth of the UES variance caused by IBS is no longer proportional to the slice current. Different from the Gaussian case in Fig.~\ref{fig:gaubench}, the result of \textsc{elegant} deviates from RF-Track even at $h=0$. This may be because each slice in \textsc{elegant}, which is cut along the current $z$, is described by a single set of second moments, so that the overlap of the mismatched slices within it cannot be resolved. For a 6D Gaussian bunch without slice mismatch, \textsc{elegant} agrees better with the other methods, as shown at $h=0$ in Fig.~\ref{fig:gaubench}. For $h=10$~m$^{-1}$ (bottom), \textsc{elegant} underestimates the effect of IBS, similar to the Gaussian case in Fig.~\ref{fig:gaubench}, whereas the slice model and RF-Track still agree well overall. 
The local differences may come from the numerical noise of RF-Track and the deviation of the transverse distribution of each slice from a Gaussian one.

\section{conclusion}

In this study, a slice model is proposed to calculate the effect of IBS on non-ideal bunches. The model is aimed at bunches with a non-Gaussian longitudinal profile, an energy chirp, and a transverse mismatch that varies from slice to slice, which are typical properties of the beams in linacs. In this paper, only the longitudinal IBS effect, i.e., the growth of the UES is considered, but the model can also be extended to the transverse directions.

For an ideal Gaussian bunch, the slice model in the frozen Gaussian approximation reduces to the conventional B\&M model. Without local over-compression, it agrees well with RF-Track after the convergence tests, and for larger chirps, it still provides a useful reference. For a more realistic bunch with an energy chirp, a transverse mismatch that varies from slice to slice, and a non-Gaussian transverse distribution, the slice model also agrees well with RF-Track, although the transverse distribution of each slice is approximated as Gaussian in the slice model. This supports the validity of the model and indirectly explains the inconsistency between the results of the two commonly used codes.

Furthermore, to treat non-ideal cases more accurately, the error caused by the non-Gaussian transverse distribution within each slice can be corrected with a Hermite--Gaussian expansion, which has recently been applied to IBS calculations~\cite{Wu2026IPAC}. For PIC codes, increasing the number of mesh points indeed makes the results converge toward the theoretical prediction, but it may further amplify the statistical error. Moreover, since the $x$--$z$ tilt changes along the beamline, each slice occupies different numbers of cells at different $s$, which may cause additional interpolation errors. Thus, a PIC code with a tilted and dynamic mesh would be beneficial for such cases. Such work is in progress, and the preliminary results show better agreement with the theory, e.g., a rise of the UES at large $h$ similar to that in Fig.~\ref{fig:gaubench}; however, their correctness remains to be verified, and the details will be reported elsewhere.

\begin{acknowledgments}

Deepseek v4.1 assisted with literature checking, mathematical exposition especially in sec.\ref{sec:III}, and manuscript preparation. This work was supported by the National Natural Science Foundation of China (12125508, 12541503, 12241501, 12205356, 12505172), the National Key Research and Development Program of China (2024YFA1612104), Shanghai Pilot Program for Basic Research – Chinese Academy of Sciences, and the International Cooperation Program, Shanghai Municipal Commission of Science and Technology (26140780400), the Shanghai Action Plan for Science, Technology and Innovation (No.24JD1402800).
\end{acknowledgments}

\appendix
\section{IBS in a straight section without dispersion}
\label{app:straight}
In this section, IBS in a straight section without transverse--longitudinal coupling is considered, in which no particles are exchanged between slices, and the slice model reduces to the conventional slice model. In such a beamline, $R_{51}=R_{52}=R_{53}=R_{54}=0$ and $R_{16}=R_{26}=0$, so that the longitudinal position $z=\kappa z_0+R_{56}\eta_0$ does not depend on the transverse plane (at low energy, $R_{56}$ caused by the velocity slippage cannot be neglected), and the centroids of each slice have only the longitudinal components, $\bm d_r=(0,0,\kappa)^{\T}$ and $\bm d_P=(0,0,h)^{\T}$, where $\kappa=R_{55}+hR_{56}$. Since the deviations are not correlated, $\Sigma_{\nu r,jj}$ and $H_{jj}$ are block diagonal, and Eqs.~\eq{alphaC} and \eq{33} give
\begin{equation}
\alpha_{jj}=\frac{\kappa^2}{2[\Sigma_{\nu,j}]_{55}},\qquad
\bm b_{jj}=\Bigl(0,\,0,\,\beta_0c\Bigl(h-\kappa\frac{[\Sigma_{\nu,j}]_{56}}{[\Sigma_{\nu,j}]_{55}}\Bigr)\Bigr)^{\T},
\label{eq:A0}
\end{equation}
 When the slice width $L$ is much larger than this width, $\omega_{ij}$ in Eq.~\eq{41} is nearly constant within the range of $P_{ij}$, i.e., $\omega_{jj}\approx1/L$ and $\omega_{ij}\approx0$ for $i\neq j$, so that only the slice itself contributes, and
\begin{equation}
A_{ij}\approx\delta_{ij}\,\frac{C_{jj}}{L}\int e^{-\alpha_{jj}\Delta^2/2}\,\Jzz^{jj}(\Delta)\,\dd\Delta.
\label{eq:A1}
\end{equation}
By the integration over $\Delta$, the average of $G_3(\bm u-\bm b_{jj}\Delta;\Sigma_{u,jj})$ with $e^{-\alpha_{jj}\Delta^2/2}$ gives a zero-mean Gaussian with the covariance $\Sigma_{u,jj}+\bm b_{jj}\bm b_{jj}^{\T}/\alpha_{jj}$, and Eq.~\eq{A1} becomes
\begin{equation}
A_{jj}\approx\frac{1}{4\pi\sigma_{x,j}\sigma_{y,j}}\,\frac{1}{\kappa L}\,
\Jzz\bigl(0;\Sigma_{u,jj}+\bm b_{jj}\bm b_{jj}^{\T}/\alpha_{jj}\bigr),
\label{eq:A2}
\end{equation}
where $\sigma_{x,j}^2=[\Sigma_{\nu,j}]_{11}$ and $\sigma_{y,j}^2=[\Sigma_{\nu,j}]_{33}$. Here, $\kappa L$ is the current length of slice $j$, so that $1/(4\pi\sigma_{x,j}\sigma_{y,j}\kappa L)$ is the position overlap factor of two particles of slice $j$, and $\Sigma_{u,jj}+\bm b_{jj}\bm b_{jj}^{\T}/\alpha_{jj}$ is the spread of the relative velocity of the particles of slice $j$ meeting at the same position; thus, by substituting Eq.~\eq{A2} into Eq.~\eq{2}, the conventional slice model is obtained, in which each slice is described by its local current, its own transverse sizes, and its local UES.

To compare with the slice formulas of Refs.~\cite{Huang2002,Lucas2026}, the UES growth rates $d\sigma_\gamma^2/ds$ given by Eqs.~\eq{2} and \eq{A2}. These formulas are evaluated with the same local current $I$, transverse beam sizes, normalized emittances, and Coulomb logarithm, and are normalized by
\begin{equation}
\Gamma_0=\frac{\sqrt{\pi}}{2}\,\frac{r_e^2\,I\ln\Lambda}{e\,c\,\sigma_r\,\varepsilon_n}.
\label{eq:A3}
\end{equation}
The parameter comes from the factor of the model in Refs.~\cite{Huang2002,Lucas2026}. Where $\sigma_\gamma=\beta_0^2\gamma_0\sigma_\eta$ is the UES in units of $m_ec^2$, $\sigma_r=\sqrt{\sigma_x\sigma_y}$, and $\varepsilon_n=\sqrt{\varepsilon_{nx}\varepsilon_{ny}}$. After this normalization, the rate of Eq.~(A3) depends only on the ratios between the diagonal elements $\sigma_{u,x}^2$, $\sigma_{u,y}^2$, and $\sigma_{u,z}^2$ of $\Sigma_{u,jj}+\bm b_{jj}\bm b_{jj}^{\T}/\alpha_{jj}$.

\begin{figure}
    \centering
    \includegraphics[width=1\linewidth]{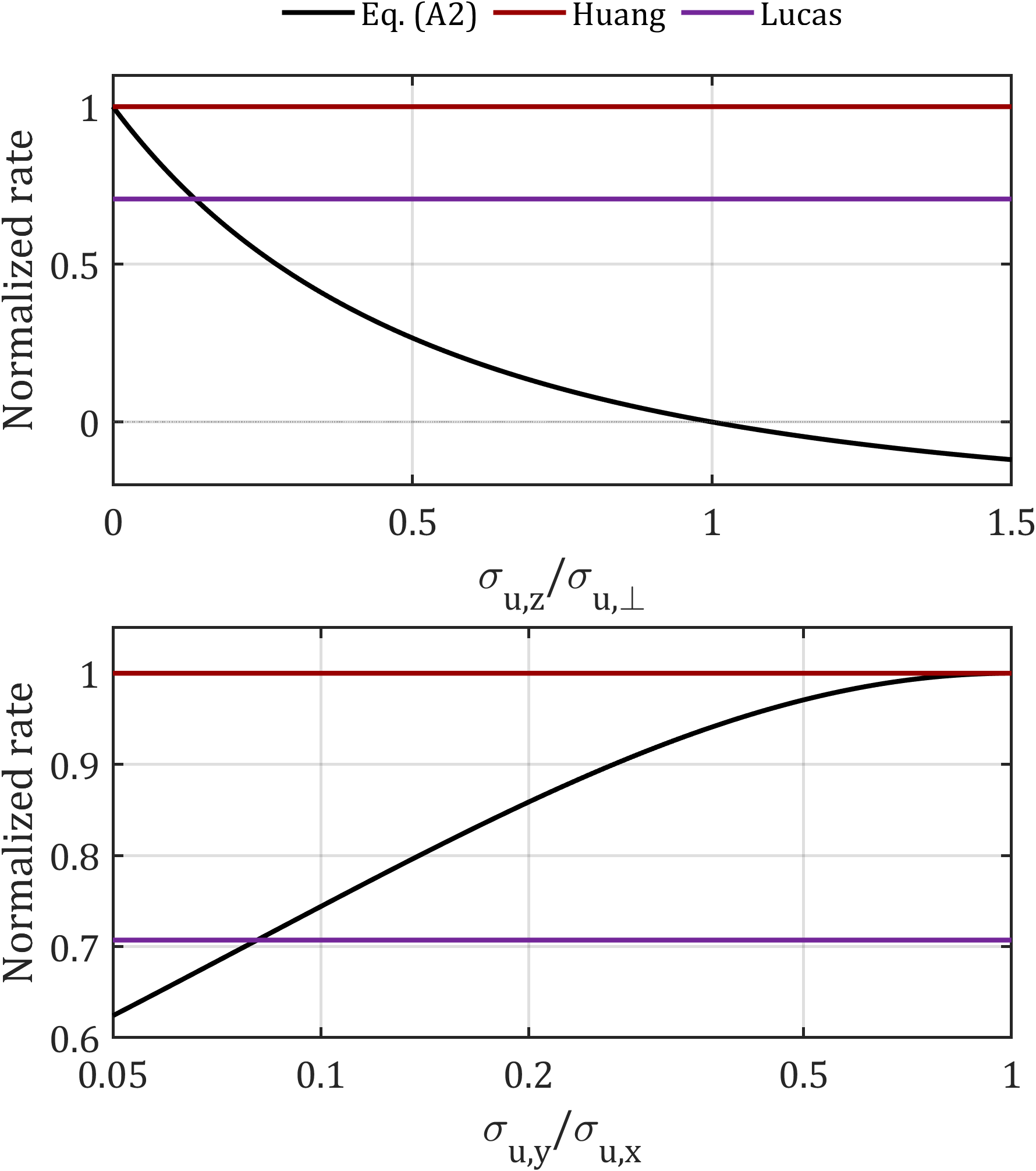}
    \caption{UES growth rates given by Eq.~(A2) and by the slice formulas of Refs.~\cite{Huang2002} and \cite{Lucas2026}, normalized by $\Gamma_0$ in Eq.~\eq{A3}, with the same local current, beam sizes, and Coulomb logarithm. (top) Round beam, $\sigma_{u,x}=\sigma_{u,y}=\sigma_{u,\perp}$, as a function of $\sigma_{u,z}/\sigma_{u,\perp}$. (bottom) Cold beam, $\sigma_{u,z}=0$, as a function of $\sigma_{u,y}/\sigma_{u,x}$.}
    \label{fig:A1}
\end{figure}

For a longitudinally cold and transversely round Gaussian bunch, Ref.~\cite{Huang2002} gives $d\sigma_\gamma^2/ds=r_e^2N_b\ln\Lambda/(4\sigma_x\varepsilon_n\sigma_z)$, where $N_b$ is the number of particles, $\sigma_z$ is the rms bunch length, and $N_b/\sigma_z=2\sqrt{\pi}\langle\lambda\rangle$ is proportional to the line density averaged over the particles, $\langle\lambda\rangle=N_b/(2\sqrt{\pi}\sigma_z)$. In Ref.~\cite{Lucas2026}, this formula is applied to a slice modeled as a Gaussian beamlet whose peak current is the slice current $I$, i.e., $N_b/\sigma_z=\sqrt{2\pi}\,I/(ec)$, so that $\langle\lambda\rangle=I/(\sqrt{2}\,ec)$ is the line density averaged over the particles of the beamlet rather than the local line density $I/(ec)$. In Fig.~\ref{fig:A1}, the normalized rates of Refs.~\cite{Huang2002,Lucas2026} are constant, since both formulas assume a longitudinally cold and transversely round beam, and that of Ref.~\cite{Lucas2026} is smaller by a factor of $\sqrt{2}$. Since $\sigma_{u,x}$ and $\sigma_{u,y}$ enter Eq.~(A3) separately without the round-beam approximation, Eq.~(A3) describes non-round beams directly, as shown in the bottom panel of Fig.~\ref{fig:A1}. Moreover, Eq.~(A3) does not take the longitudinally cold limit, so that the UES generated by IBS is retained in $\sigma_{u,z}$ and feeds back into the subsequent rate, as shown in the top panel of Fig.~\ref{fig:A1}; thus, Eq.~(A3) still provides a reliable estimate for beams in which the IBS-induced UES growth is significant.

\bibliography{refs}

\end{document}